\documentclass[preprint]{aastex}
\shorttitle{Activity-Sensitive Lines}
\shortauthors{Wise \& Dodson-Robinson}

\begin{document}

%% LaTeX will automatically break titles if they run longer than
%% one line. However, you may use \\ to force a line break if
%% you desire.

\title{New Methods for Finding Activity-Sensitive Spectral Lines: Combined Visual Identification and an Automated Pipeline Find a Set of 40 Activity Indicators}

%% Use \author, \affil, and the \and command to format
%% author and affiliation information.
%% Note that \email has replaced the old \authoremail command
%% from AASTeX v4.0. You can use \email to mark an email address
%% anywhere in the paper, not just in the front matter.
%% As in the title, use \\ to force line breaks.

\author{A. W. Wise\altaffilmark{1,2}, S. E. Dodson-Robinson\altaffilmark{1,3}, K. Bevenour\altaffilmark{1}, and A. Provini\altaffilmark{1}}
%\affil{Department of Physics and Astronomy, University of Delaware, Newark, DE 19711}
%\affil{Department of Astronomy, Yale University, New Haven, CT 06520}

%% Notice that each of these authors has alternate affiliations, which
%% are identified by the \altaffilmark after each name.  Specify alternate
%% affiliation information with \altaffiltext, with one command per each
%% affiliation.

\altaffiltext{1}{Department of Physics and Astronomy, University of Delaware, Newark, DE 19711}
\altaffiltext{2}{Email: aww@udel.edu}
\altaffiltext{3}{Bartol Research Institute, Newark, DE 19711}
%\altaffiltext{3}{Department of Astronomy, Yale University, New Haven, CT 06520}

%% Mark off your abstract in the ``abstract'' environment. In the manuscript
%% style, abstract will output a Received/Accepted line after the
%% title and affiliation information. No date will appear since the author
%% does not have this information. The dates will be filled in by the
%% editorial office after submission.

\begin{abstract}
Starspots, plages, and activity cycles cause radial velocity variations that can either mimic planets or hide their existence. To verify the authenticity of newly discovered planets, observers may search for periodicity in spectroscopic activity indices such as Ca H \& K and H$\alpha$, then mask out any Doppler signals that match the activity period or its harmonics. However, not every spectrograph includes Ca H \& K, and redder activity indicators are needed for planet searches around low-mass stars. Here we show how new activity indicators can be identified by correlating spectral line depths with a well-known activity index. We apply our correlation methods to archival HARPS spectra of $\epsilon$~Eri and $\alpha$~Cen~B and use the results from both stars to generate a master list of activity-sensitive lines whose core fluxes are periodic at the star's rotation period. Our newly discovered activity indicators can in turn be used as benchmarks to extend the list of known activity-sensitive lines toward the infrared or UV. With recent improvements in spectrograph illumination stabilization, wavelength calibration, and telluric correction, stellar activity is now the biggest noise source in planet searches. Our suite of $> 40$ activity-sensitive lines is a first step toward allowing planet hunters to access all the information about spots, plages, and activity cycles contained in each spectrum.
\end{abstract}


%% From the front matter, we move on to the body of the paper.
%% In the first two sections, notice the use of the natbib \citep
%% and \citet commands to identify citations.  The citations are
%% tied to the reference list via symbolic KEYs. The KEY corresponds
%% to the KEY in the \bibitem in the reference list below. We have
%% chosen the first three characters of the first author's name plus
%% the last two numeral of the year of publication as our KEY for
%% each reference.

%% Authors who wish to have the most important objects in their paper
%% linked in the electronic edition to a data center may do so by tagging
%% their objects with \objectname{} or \object{}.  Each macro takes the
%% object name as its required argument. The optional, square-bracket 
%% argument should be used in cases where the data center identification
%% differs from what is to be printed in the paper.  The text appearing 
%% in curly braces is what will appear in print in the published paper. 
%% If the object name is recognized by the data centers, it will be linked
%% in the electronic edition to the object data available at the data centers  
%%
%% Note that for sources with brackets in their names, e.g. [WEG2004] 14h-090,
%% the brackets must be escaped with backslashes when used in the first
%% square-bracket argument, for instance, \object[\[WEG2004\] 14h-090]{90}).
%%  Otherwise, LaTeX will issue an error. 

\section{Introduction}

%Precision radial velocity (RV) measurements have become a prolific method of confirming and discovering exoplanets in the past two decades (see e.g. \cite{wright17} for a review). However, stellar activity introduces noise into RV time series that can obscure the signals of low-mass planets \citep{meunier10,barnes11,lovis11,andersen15}.
Although NASA is investing tremendous resources in space missions designed to find habitable planets and detect biomarkers \citep{gardner04, ricker14}, the search for Earthlike planets orbiting Sunlike stars still requires ground-based radial velocity measurements to identify or confirm targets for space-based observations \citep[e.g.][]{mayor14}. Of the main radial velocity (RV) noise sources which obscure the signals of low-mass planets, stellar noise is one of the most intractable.
Indeed, there is concern that stellar jitter may create an intrinsic RV precision floor
%of 1~m~s$^{-1}$,
near the 0.5-1~m~s$^{-1}$ precision of current-generation RV instruments, which would limit the minimum mass of a detectable planet in the habitable zone of $\alpha$~Cen~B, a K1 dwarf, to $2.5 M_{\oplus}$ \citep[e.g.][]{dumusque11}. For next-generation RV instruments expected to reach instrumental precisions of 10-20 cm/s, stellar jitter is widely anticipated to be the dominant source of RV noise \citep{pepe13, jurgenson16}.
Yet stellar jitter is not white noise, but is instead structured in time: p-modes have timescales of minutes \citep{christensendaalsgard04,kjeldsen05}, granulation and supergranulation have timescales of hours to days \citep{derosa04,brandt08}, and magnetic activity shows timescales ranging from the stellar rotation period \citep{saar97} to multi-year cycles \citep{baliunas95}.
Methods that either suppress stellar jitter \citep{pepe02, angladaescude12} or flag it using stellar activity signals and model it out \citep{aigrain12, dumusque12, dumusque14, haywood14, rajpaul15, giguere16, meunier17} allow us to push below the stellar noise floor toward true earth analogs.

Although RV corrections for stellar activity are often based on only one or a few lines \citep[such as Ca$_{\rm II}$ H \& K, Na D, or H$\alpha$;][]{lovis11, barnes14, robertson13, robertson15}, hundreds or even thousands of spectral lines are present in most planet-search spectra. Including information from as many lines as possible is an important step toward improving activity diagnostics \citep{giguere16}. Here we present a new method of identifying all activity-sensitive lines in the HARPS (High Accuracy Radial velocity Planet Searcher) spectrograph's wavelength range of $3800 \AA - 6900 \AA$. The methods presented here are not particular to any wavelength range, so they could be replicated to find activity-sensitive lines in any wavelength range that contains a previously known stellar activity indicator. Our final product is a new list of $> 40$ activity-sensitive spectral lines whose depths vary with the same periodicity as $\log(R'_{HK})$, an index measuring activity-induced emission in the cores of the Ca$_{\rm II}$ H \& K absorption lines. Our new activity indicators can be measured using the same high-resolution spectra that are used to make precise RV measurements.

%Previously used activity indicators (wide variety, not all in visible light): Criscuoli et al 2013, Barnes et al 2014, Lorenzo-Oliveira et al 2016, Robertson et al 2013, Robertson et al 2015, Robertson et al 2016, Sasso et al 2017

The paper is organized as follows. In Section~\ref{finding}, we give a step-by-step explanation of how we derived our list of activity-sensitive lines. In Section~\ref{additional}, we describe the tests we performed to validate our criteria for labeling these lines as activity sensitive. Finally, in Section~\ref{discussion}, we discuss how our results might be used to improve RV techniques for finding exoplanets.

\section{Finding Activity-Sensitive Lines}\label{finding}

%As we look for regions of the spectrum that can be used as new stellar activity indicators, we focus on spectral absorption lines, as opposed to other arbitrary wavelength intervals, because we expect absorption lines' physical underpinning to allow activity-sensitive lines found with one star to be valid tracers of activity on other stars of similar temperature. However, the following methodology could be used to probe stellar activity in any set of wavelength intervals that need not contain absorption lines.

In our study of activity-sensitive spectral lines, we have two goals: (1) to identify lines aside from the frequently used Ca$_{\rm II}$ H \& K, H$\alpha$, and Na D that can add information about each star's chromospheric and/or photospheric activity, and (2) to understand the physical reasons why certain lines are especially sensitive to activity. Rather than simply relying on a machine-learning algorithm, we use a two-pronged approach that helps us build our physical intuition about how activity affects line depths.

Our dataset consists of spectra from the HARPS echelle spectrograph operating in High Accuracy Mode (HAM), at the ESO La Silla 3.6m telescope \citep{mayor03}. We use the wavelength-calibrated, order-by-order extracted 2D spectra (e2ds) provided by the HARPS Data Reduction Software (DRS) pipeline version 3.5. The DRS pipeline also provides extracted, wavelength-calibrated 1-D spectra along with cross-correlation functions and bisector spans, which we do not use in this work. We obtain the e2ds spectra along with associated wavelength and blaze correction files from the public ESO archive.

We analyze HARPS spectra from two stars: $\epsilon$~Eri and $\alpha$~Cen~B. We chose these stars because they are both nearby and well-known, which means they have abundant spectra in the HARPS archive, and because they have similar spectral types (K2 and K1 respectively), so we can compare their activity-sensitive spectral features without considering effects of stellar temperature. We also chose these two stars because $\epsilon$~Eri is very active ($\log{R'_{HK}} = -4.46$; \citealp{brandenburg17}) and $\alpha$~Cen~B is relatively inactive ($\log{R'_{HK}} = -4.93$), so we can probe a wide range in stellar activity levels with just these two stars. Table~\ref{properties} provides a summary of the properties of each star. In \S~\ref{EpsEri} we describe our visual analysis of the correlations between line depth and activity in $\epsilon$~Eri, and in \S~\ref{AlphaCenB} we discuss our automated procedure for identifying activity-sensitive lines in $\alpha$~Cen~B.

%We analyze spectra of two stars: $\epsilon$~Eri, a K2V star known for being unusually magnetically active, and $\alpha$~Cen~B, a quiet K1V star representative of terrestrial planet search targets.

%(MAYBE CUT THIS NEXT PART - IT IS HARD TO SUMMARIZE PROCEDURE CLEARLY YET BRIEFLY) Our procedure is as follows: $\epsilon$~Eri spectra are used to generate plots of pixel-by-pixel activity for all 72 orders of the HARPS spectra. These plots are used to visually identify lines of interest. Line depth and area below half-depth flux are measured for each line of interest in the $\alpha$~Cen~B spectra, and these line properties are correlated against S-index and used to generate Lomb-Scargle Periodograms. Activity sensitive lines are chosen based on strong correlations with S-index, and periodogram peaks near the stellar rotation period.

\begin{table}
%\centering
\caption{Stellar Properties\label{properties}}
\begin{tabular}{lll}
\tableline\tableline
Property & $\epsilon$~Eri & $\alpha$~Cen~B\\
\tableline
Spectral type & K2\tablenotemark{2} & K1\tablenotemark{2}\\
Effective temperature (K) & 5152\tablenotemark{2} & 5230\tablenotemark{2}\\
Metallicity [Fe/H] & 0.0\tablenotemark{2} & 0.27\tablenotemark{2}\\
$\log{g}~(\mathrm{cm~s^{-1}})$ & 4.57\tablenotemark{3} & 4.37\tablenotemark{4}\\
$\log{R'_{HK}}$ & -4.46\tablenotemark{2} & -4.93\tablenotemark{2}\\
Rotation period (days) & 11.1\tablenotemark{2} & 36.2\tablenotemark{2}\\
Age (Gyr) & 0.6\tablenotemark{2} & 5.4\tablenotemark{2}\\
Mass (M$_\sun$) & 0.82\tablenotemark{1} & 0.907\tablenotemark{5}\\
\tableline
\end{tabular}
%\tablenotetext{a}{See \cite{ss73}}
\tablerefs{(1)~\citealt{baines12}; (2)~\citealt{brandenburg17}; (3)~\citealt{clem04}; (4)~\citealt{gilli06}; (5)~\citealt{thevenin02}.}
\end{table}

\subsection{Visual exploration of line-depth variation}\label{EpsEri}

To visually identify activity-sensitive lines in plots, we use archival HARPS spectra of the magnetically active star $\epsilon$~Eri. Our goal is to identify which, if any, of the 4096 resolution elements (henceforth referred to as pixels) in each echelle order co-vary with the Mt.\ Wilson S-index, a robust tracer of stellar activity \citep{wilson78}. We obtained 557 $\epsilon$~Eri spectra from the ESO archive and threw out 17 spectra due to either outlying S-index measurements or poor line-depth measurements\footnote{This was accomplished visually using our S-index correlation plots (see Section~\ref{AlphaCenB}); a complete list of spectra used is available upon request.}. The 540 remaining spectra have observation dates ranging from November 2003 to August 2015, allowing us to capture most of the star's $\sim$13 year activity cycle identified by \cite{metcalfe13}. However, these 540 spectra were taken on 30 different nights, and 422 of them were taken on two consecutive nights: October 25-26, 2004. Because of this uneven temporal sampling of the $\epsilon$~Eri spectra, in the following analysis we do not use the time at which each spectrum was taken, but instead look at how the spectra change as a function of S-index.

\begin{figure}
\epsscale{1.0}
\plotone{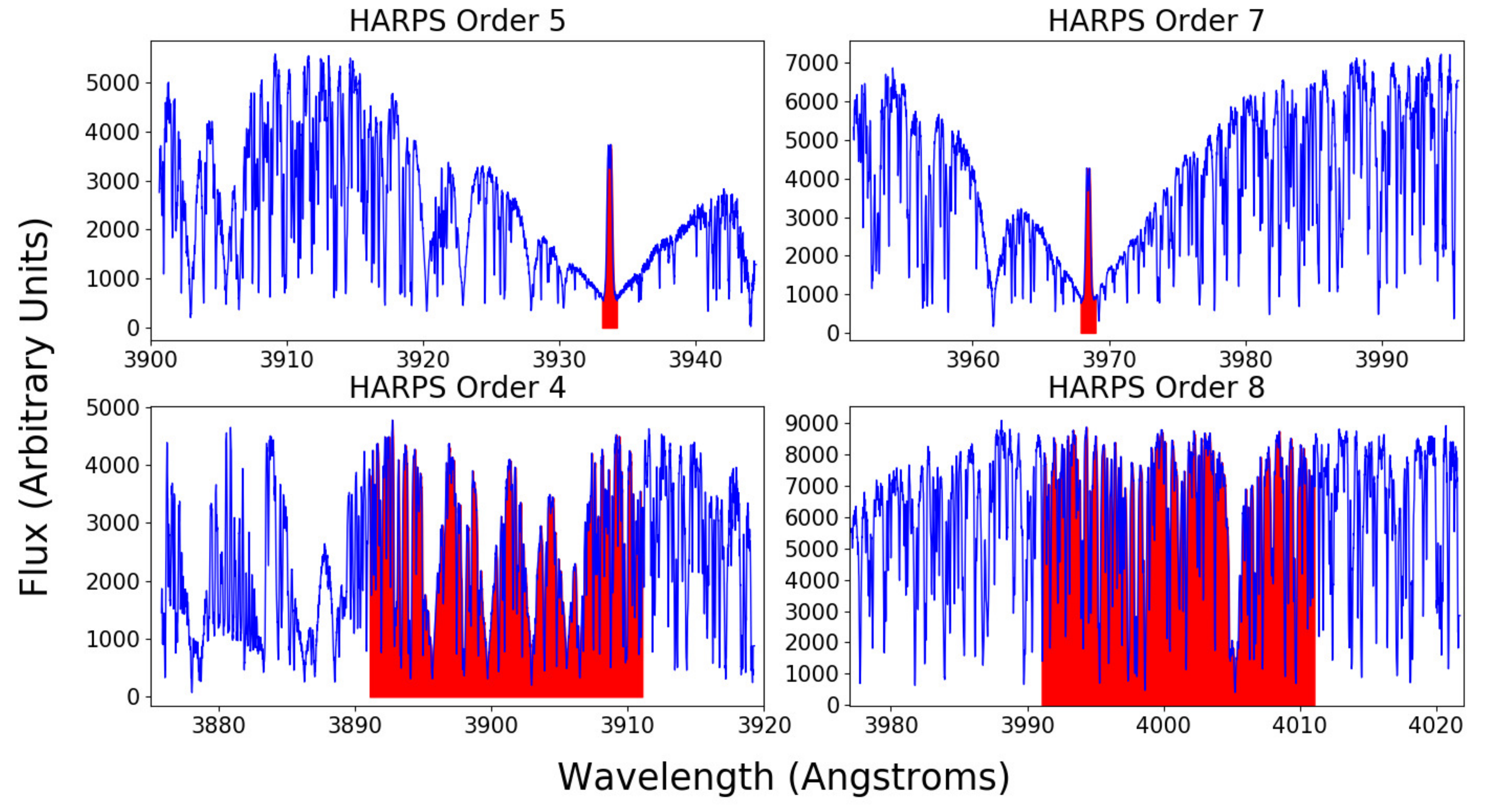}
\caption{Example measurement of the Mt.\ Wilson S-index of a HARPS spectrum of $\epsilon$~Eri. Each order is blaze corrected using calibrations provided by the HARPS team. The S-index is calculated by adding together the wavelength-integrated flux in the calcium H and K line cores (top), and dividing the result by the sum of the wavelength-integrated flux in the two continuum regions (bottom). See \cite{lovis11} for specific wavelength intervals. \label{fig1}}
\end{figure}

Each $\epsilon$~Eri spectrum is RV-shifted to the frame of the star using the difference between the barycentric earth radial velocity and barycentric object radial velocity, both taken from FITS (Flexible Image Transport System) file headers. Each spectrum is then divided by its blaze function estimate (also obtained from the ESO archive), and aligned onto a common wavelength grid using a linear interpolation. After calculating an S-index for each spectrum (as in \cite{lovis11}, see Figure~\ref{fig1}), we place the spectra in bins of equally sized ranges in S-index, under the constraint that each bin contains at least ten spectra, giving us eight bins between S = 0.388 and S = 0.469 for this dataset. Since the blaze-corrected spectra are not yet normalized, we correct for flux variations due to variable exposure time and inter-order blaze variability by dividing each order of each spectrum by its trimmed mean flux (average of 5th to 95th percentile of all 4096 pixels in the order). Finally, for each wavelength data point in a given S-index bin, we take a trimmed mean (average of values between the 25th and 75th percentile) of the normalized fluxes at that wavelength in the S-index bin. The end result is eight trimmed mean spectra, one for each S-index bin, that represent the star during different phases of the activity cycle.

To look for variations among these trimmed mean spectra, we calculate the absolute value of A-Q, where Q is the flux in the spectrum from the lowest S-index bin (quietest), and A is the flux from any of the other seven (more active) trimmed-mean spectra. Finally, we make plots like Figure~\ref{fig2}, where the seven more active trimmed mean spectra are stacked vertically, and each data point is colored according to its $|({\rm A-Q})|$ value. In Figure~\ref{fig2}, we have made higher $|({\rm A-Q})|$ values darker, so activity-sensitive lines are those that are light-colored at the top of the plot (low S-index) and gradually get darker toward the bottom of the plot (high S-index). In this particular spectral order we see strong activity indicators at $4851 \AA$, $4861 \AA$, and $4865 \AA$. To avoid false positives, we make sure lines of interest are away from the noisy edges of orders, where pixel variability may simply be due to Poisson noise. We use an interactive \texttt{ipython} program that allows us to quickly record the start and end wavelengths of each line we visually identify.

\begin{figure}
\epsscale{1.0}
\plotone{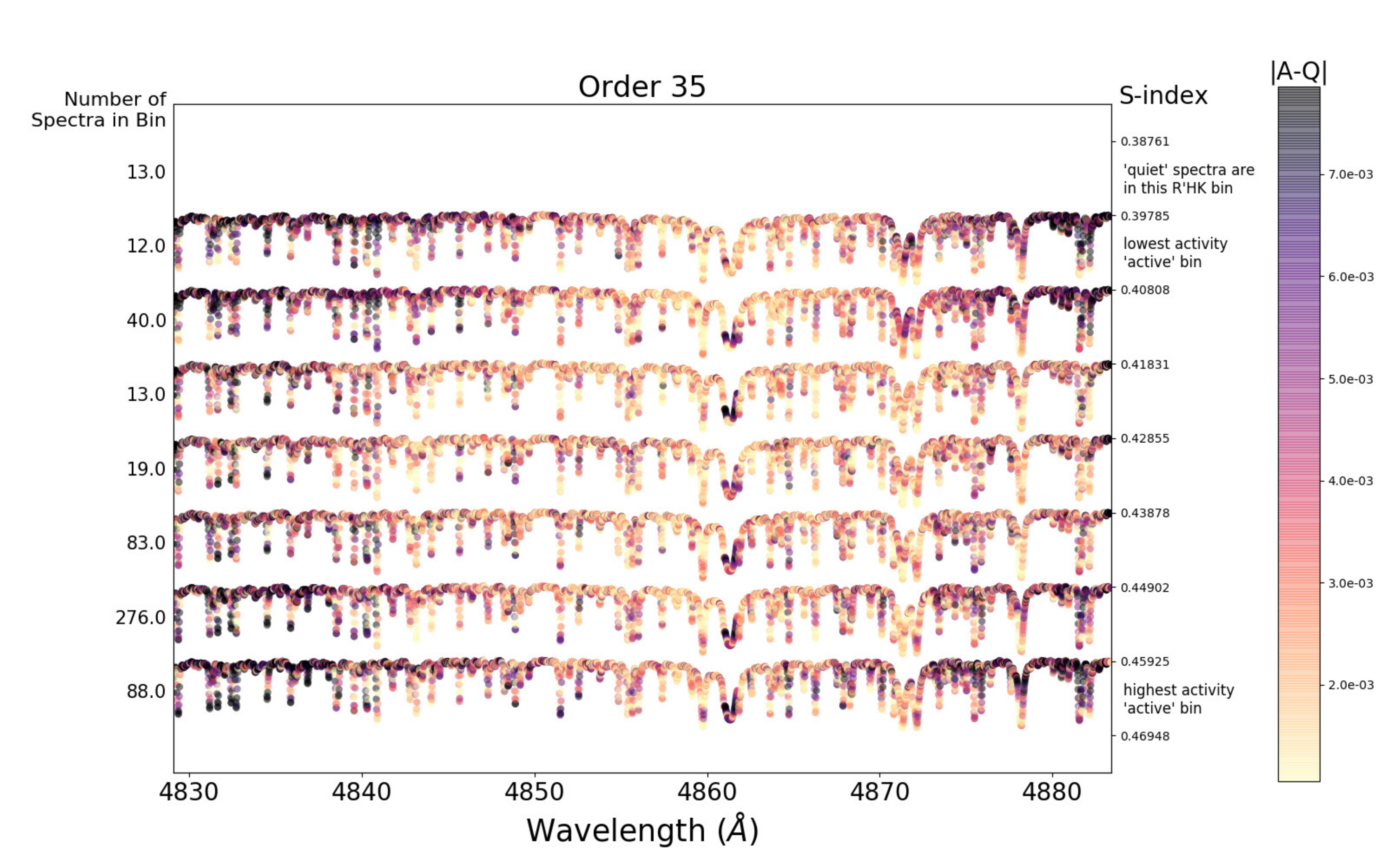}
\caption{Visualization of a single order of pixel-by-pixel activity for HARPS spectra of $\epsilon$~Eri.
Since higher values of $|A-Q|$ are darker and we expect more activity in spectra with higher S-index, we look for lines that go from light to dark from the top to the bottom of the plot.
The full set of 72 pixel-by-pixel activity plots (one per order) is available upon request. \label{fig2}}
\end{figure}

\iffalse

\figsetstart
\figsetnum{2}
\figsettitle{Pixel-by-Pixel Activity Levels\label{fig2}}
 
\figsetgrpstart
\figsetgrpnum{2.1}
\figsetgrptitle{Order 1}
\figsetplot{f2.eps}
\figsetgrpnote{image 1 caption}
\figsetgrpend

\figsetgrpstart
\figsetgrpnum{2.2}
\figsetgrptitle{Order 2}
\figsetplot{f2.eps}
\figsetgrpnote{image 1 caption}
\figsetgrpend

\figsetgrpstart
\figsetgrpnum{2.3}
\figsetgrptitle{Order 3}
\figsetplot{f2.eps}
\figsetgrpnote{image 1 caption}
\figsetgrpend
 
\figsetend

\fi

%\clearpage

By using plots like Figure~\ref{fig2} for nearly all of the 72 orders in HARPS spectra (orders with a very low SNR or heavy telluric contamination are excluded), we visually identified 193 lines of interest. To verify the robustness of our visually identified activity indicators, and to ensure that the lines' core fluxes correlate with the Mt.\ Wilson S-index even for a quieter star, we performed a similar, but fully automated calculation using spectra from $\alpha$~Cen~B.

\subsection{Automated verification of activity-sensitive lines}\label{AlphaCenB}

RV survey targets tend to have much weaker stellar activity then $\epsilon$~Eri. To make sure our lines of interest have measurable activity signals in a less active star, we filter our list of lines of interest from the visual analysis by looking for correlations between line properties and S-index in a nearby low-activity star, $\alpha$~Cen~B.

\begin{figure}
\epsscale{1.0}
\centering
%\plotone{fig-plotHalfDepths.png}
\plotone{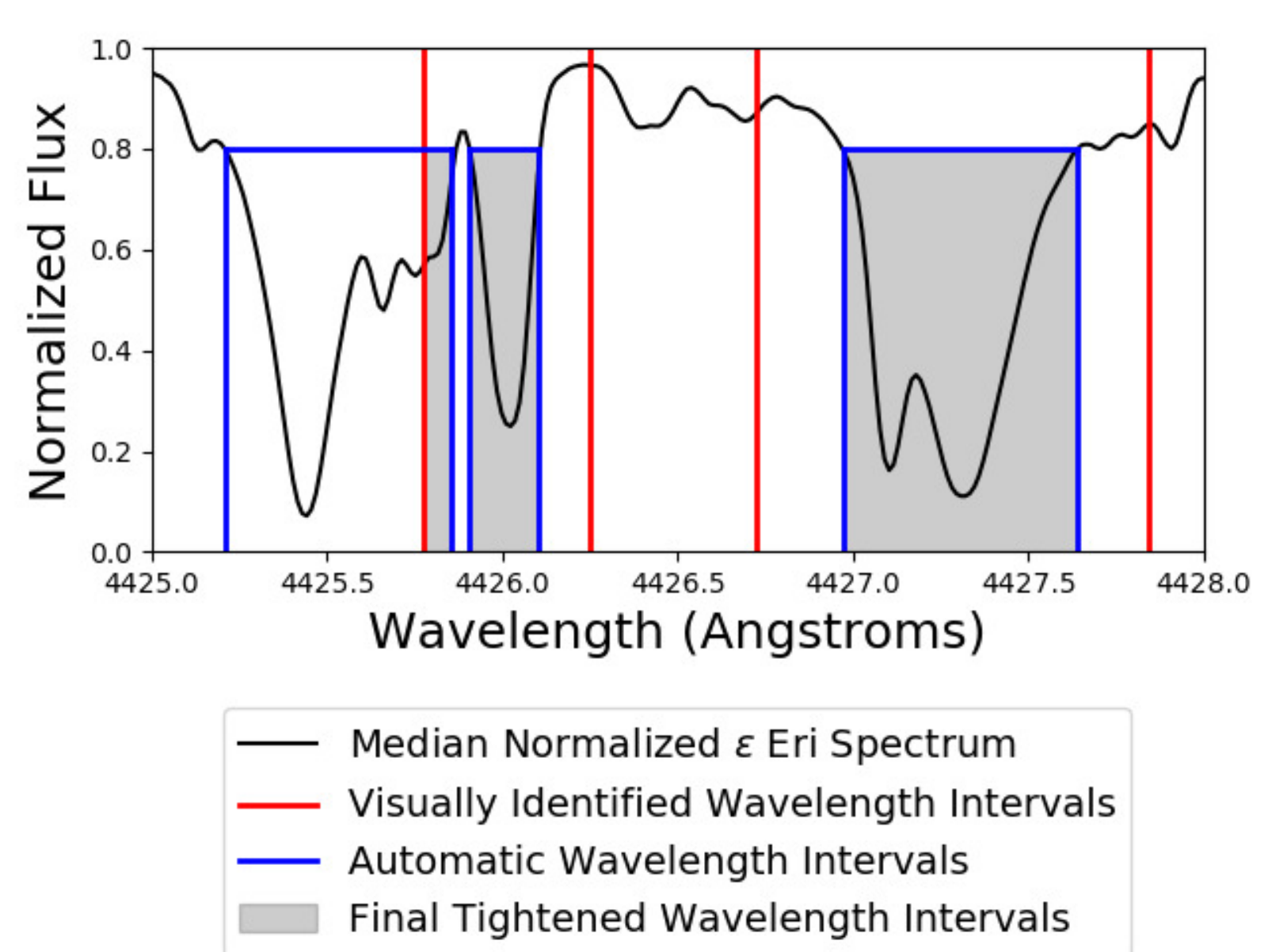}
\caption{Cartoon showing how we automatically tightened our visually identified wavelength intervals by taking the intersection of an automatically generated list of line wavelength intervals with our visual intervals. The line at 4426~$\AA$ was correctly separated from a line to its left that was within its visually identified wavelength interval. The line at 4425.4~$\AA$ did not make it into our final tightened wavelength intervals because it showed no indication of activity in plots like Figure~\ref{fig2}. Some lines are too closely blended to be separated, such as in the wavelength interval on the right, so our automated line property measurements are really measuring properties of the blend. \label{fig3}}
\end{figure}

Before we measure line properties in $\alpha$~Cen~B spectra, we take the following steps to tighten our wavelength intervals around each visually-identified line in the $\epsilon$~Eri spectra, preventing other nearby lines from influencing line property measurements. This time we continuum normalize all of the $\epsilon$~Eri spectra using the procedure in Appendix~\ref{norm}, in between the steps of dividing out the blaze functions and aligning the spectra (as described in Section~\ref{EpsEri}). Using these 540 continuum normalized, wavelength aligned spectra, we find the median normalized flux at each pixel. We use a custom algorithm to identify $\sim$1500 lines in the median spectrum (for a description of the algorithm, see Section~\ref{additional}), and take the intersection of this set of wavelength intervals with the set of wavelength intervals for activity-sensitive lines that we recorded during our visual analysis in Section~\ref{EpsEri}.
This gives us a tightly fitting wavelength interval around each visually-identified line, as shown in Figure~\ref{fig3}.
Finally, using the median spectrum we compute the normalized flux at half of each line's maximum depth. We label this metric, which is a value between 0 and 1 assigned to each line, ``half-depth flux.'' Now that we have our tightly fitting wavelength intervals and associated half-depth fluxes from the $\epsilon$~Eri spectra, we can move toward measuring line properties on $\alpha$~Cen~B spectra.

We obtained 2549 $\alpha$~Cen~B spectra spanning 23 March 2010 -- 12 June 2010 from the ESO archive. We selected this time period because of its high cadence of observations (spectra on 50 out of 82 nights) and rotationally modulated $\log{R'_{HK}}$ signal \citep{dumusque12}. \cite{thompson17} realized that this was an ideal dataset to study stellar magnetic activity in K-dwarfs, and presented evidence that one can use it to find a large number of spectral lines whose depth and radial velocities covary with $\log R^{'}_{HK}$. Each of these spectra is actually a short sub-exposure from one of the 1--3 full exposures taken on a given night. We treat each spectrum as an individual measurement and do not bin the $\alpha$~Cen~B spectra. Out of the 2549 spectra we obtained, we threw out 58 due to visually outlying line-depth measurements (as seen in correlation plots at the end of this section).

\begin{figure}
\epsscale{1.0}
\centering
\plotone{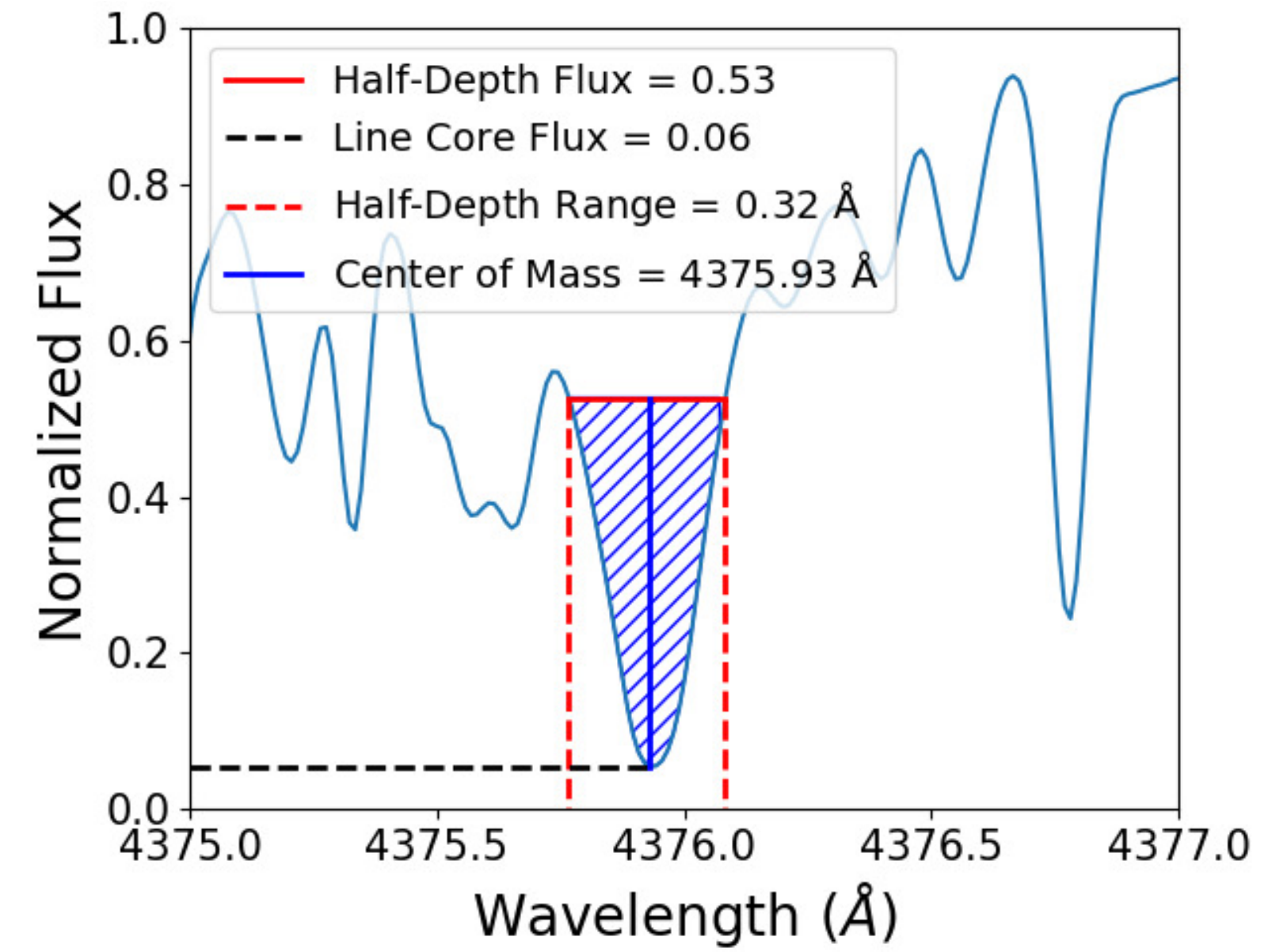}
\caption{Cartoon showing how we calculated each of our three line measurements (core flux, half-depth flux, and center of mass) for each visually identified line in each spectrum. The blue diagonally hatched area is the integral area for the center of mass calculation. The value of half-depth flux, which is only calculated once for each line on the median spectrum from the entire $\epsilon$~Eri dataset, is shown for reference.\label{fig4}}
\end{figure}

We RV-shift each $\alpha$~Cen~B spectrum to the frame of the star and remove the blaze using the RVs and blaze corrections provided on the ESO archive, just as we did for the $\epsilon$~Eri spectra. However, this time we do not interpolate onto a common wavelength grid, as this is unnecessary to measure line properties in individual spectra, and we found it would introduce 10-20\% scatter into our measurements of line properties. Next the $\alpha$~Cen~B spectra are continuum normalized using the procedure in Appendix~\ref{norm}. Finally, we use a cubic spline interpolation to fit each order of each spectrum, and perform the following measurements on the cubic spline function in each line's wavelength interval (see Figure~\ref{fig4}):

\begin{itemize}
\item Line core flux: absolute minimum of the cubic spline. Line core flux is similar to the $I_{{\rm H}\alpha}$ index developed by \citet{kurster03} to measure ``filling-in'' of the H$\alpha$ line due to chromospheric activity.
\item Half-depth range: difference between the two wavelengths where the spline equals the line half-depth flux. A line that Zeeman-splits in starspot magnetic fields, such as Fe $5250 \AA$, would have large half-depth range when starspots appeared.
\item Center of mass: $\int \lambda \times (F_{\rm hd} - S) d\lambda\ /  \int (F_{\rm hd} - S ) d\lambda$ over the half-depth range wavelengths, where $F_{\rm hd}$ is the half-depth flux, $S$ is the cubic spline, and $\lambda$ is the wavelength. Lines strongly affected by convective inhibition would Doppler shift as active regions rotated in and out of view, causing a change in center of mass.
\end{itemize}

\begin{figure}
\centering
\includegraphics[scale=0.45]{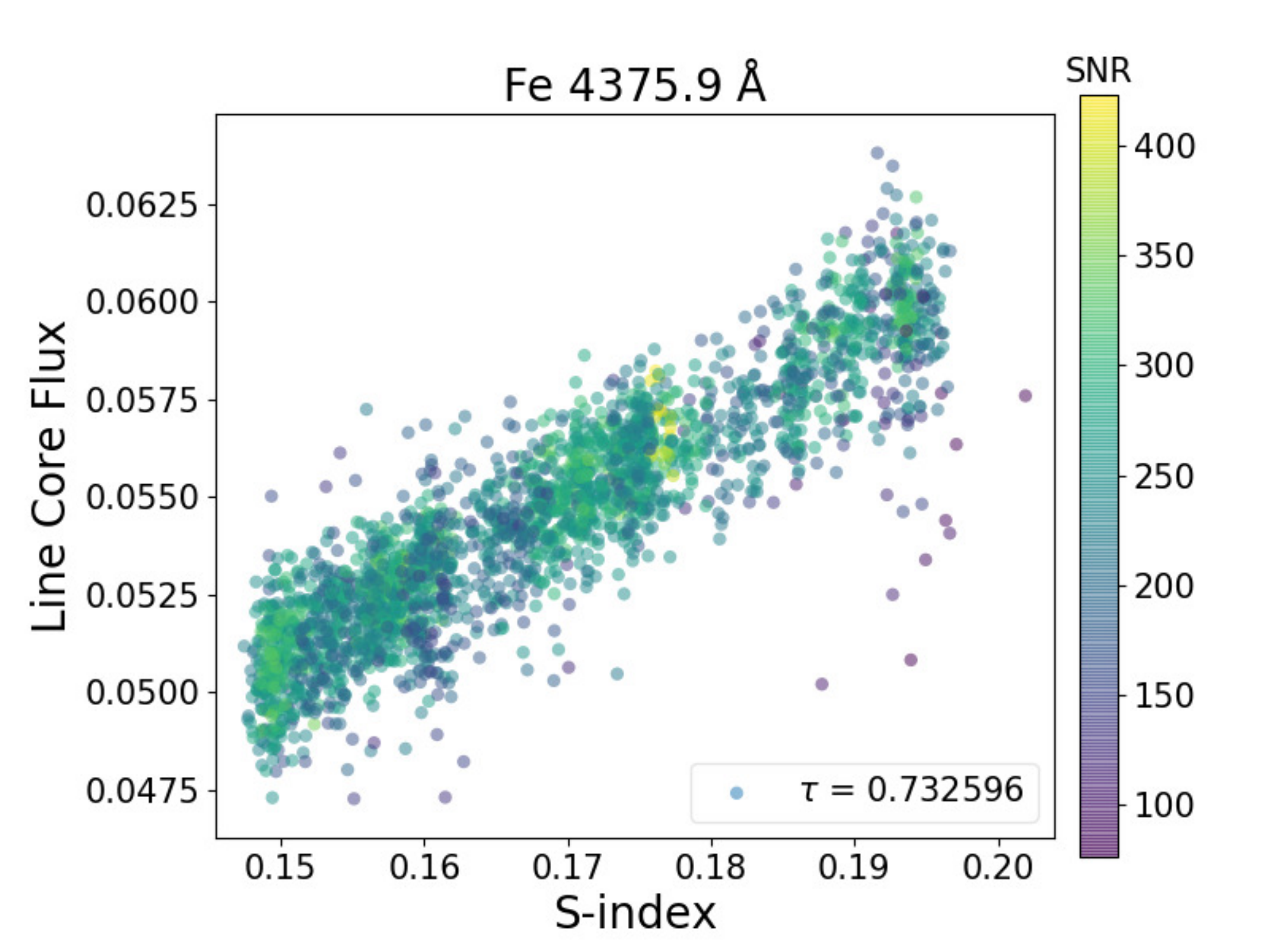}
\includegraphics[scale=0.45]{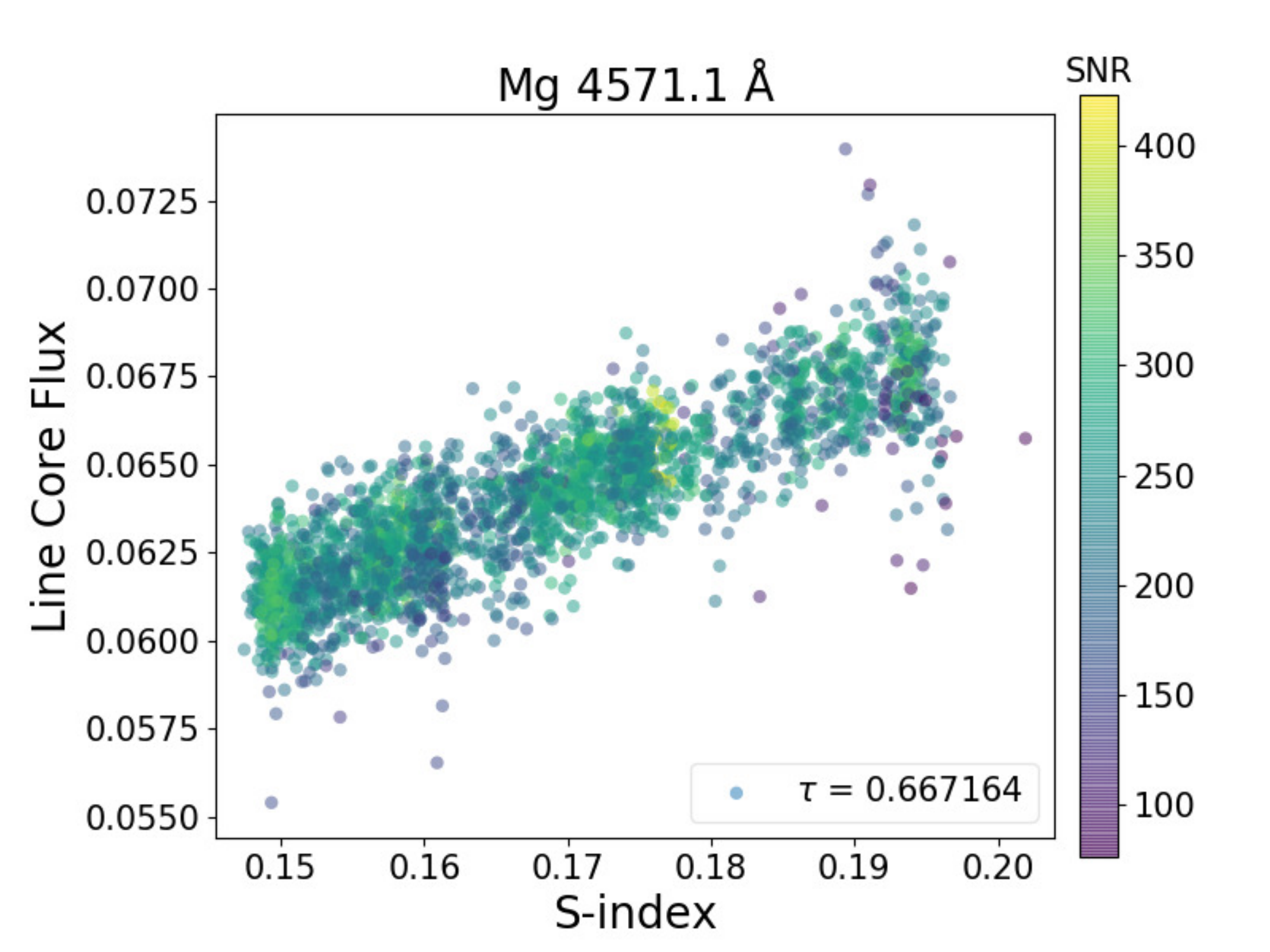}
\includegraphics[scale=0.45]{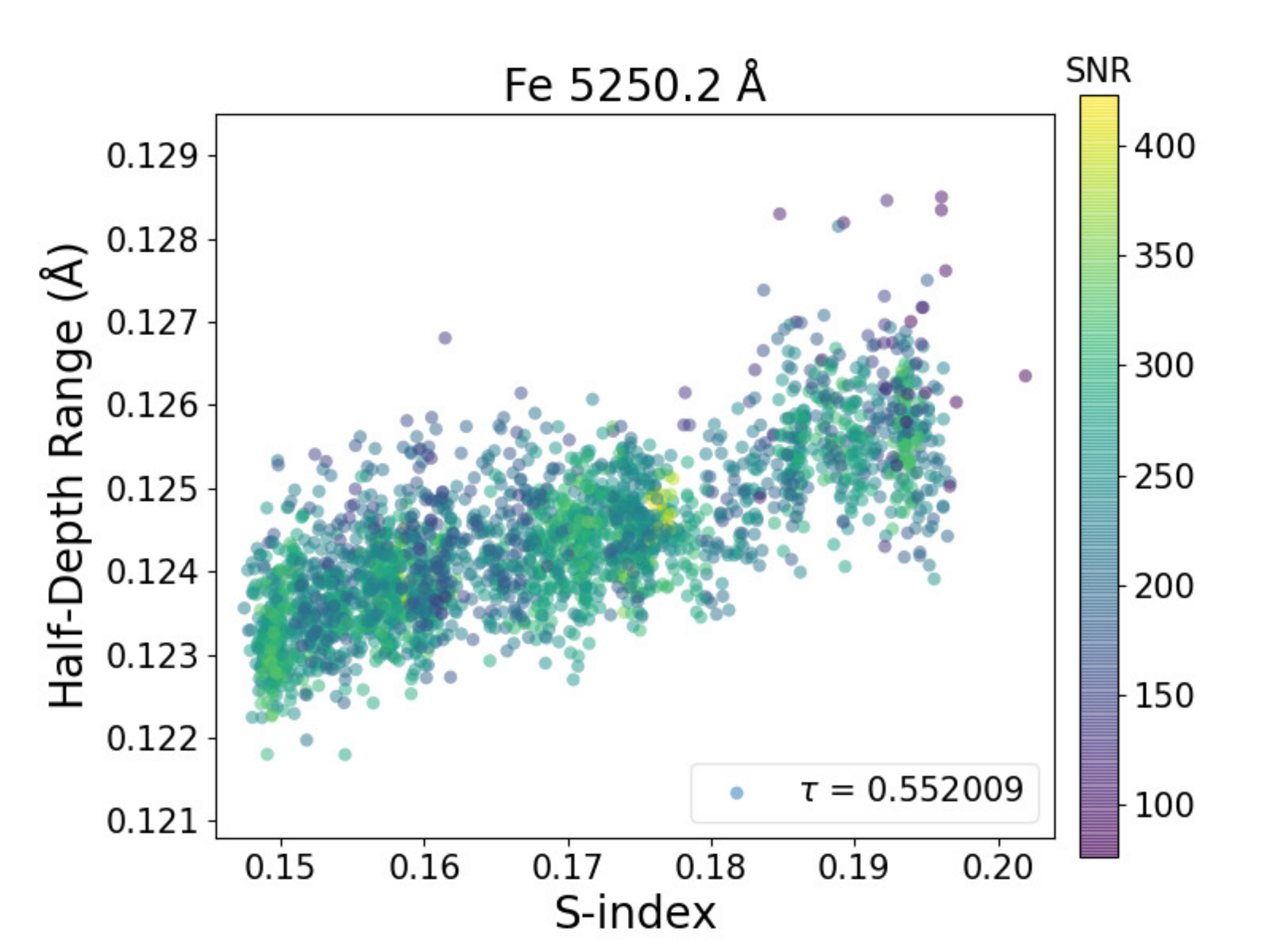}
\caption{Plots showing correlations between selected line properties and S-index for lines at $4375.9 \AA$, $4571.1 \AA$, and $5250.2 \AA$. Each point, a line core flux or half-depth range measured from a single spectrum, is colored according to signal-to-noise ratio.\label{fig5}}
\end{figure}

We also calculate the S-index of each $\alpha$~Cen~B spectrum using the same procedure as in Section~\ref{EpsEri}. For each line of interest from Section~\ref{EpsEri}, we calculate a Kendall's Tau coefficient, $\tau$, for the correlation between the S-index measurements and each of these three line properties. Example correlation plots for three lines are shown in Figure~\ref{fig5}. We chose the correlation coefficient as our metric for sensitivity to activity because it captures both the covariance of one of our measurables (line core flux, half-depth range, or center of mass) with S-index, and the noise intrinsic to measurement. In terms of the correlation plots in Figure~\ref{fig5}, the covariance of line core flux with activity is proportional to the slope the best-fit line, and the intrinsic noise is represented by the scatter in line core flux within a small window in S-index.

%This method could also be used where such a periodicity has not been identified, by stacking the periodograms from activity sensitive lines (identified as in the previous section) vertically and finding the highest peak.

We pick out all lines of interest with $| \tau | \geq 0.5$ for at least one of the three measured line properties, and add these to our list of activity-sensitive lines. We chose 0.5 for the cutoff because it is the value above which our visual line search, based on plots like Figure~\ref{fig2}, finds nearly all of the same lines as our automated procedure (see completeness analysis in Section~\ref{additional}). Of the 193 activity-sensitive lines we identified using plots like Figure~\ref{fig2} of the $\epsilon$~Eri spectra, 39 had line core flux correlation coefficients of $| \tau | \geq 0.5$. 3 of these 39 lines also had half-depth range coefficients $| \tau | \geq 0.5$, and 4 of the remaining 154 lines that did not meet our $| \tau | \geq 0.5$ cutoff for line core flux did have $| \tau | \geq 0.5$ for half-depth range. After finding that line core flux was sufficient to determine line-list membership for all of our activity-sensitive lines except four,
%Fe $5250 \AA$ which, as expected, was identified by half-depth range,
%\footnote{At $5250.2 \AA$, line core flux correlation was 0.472 and area below half-depth flux correlation was 0.548}
we modified our algorithm to only use line core flux to determine list membership. We suspect these four lines, $4626.2 \AA$, $5250.2 \AA$, $5890.0 \AA$, and $6141.7 \AA$ may probe photospheric effects such as Zeeman splitting in starspots. However, our results suggest that core flux changes such as line-core infilling from chromospheric emission or changing absorption probability are the strongest signals out of the various ways stellar activity can affect spectral lines.

Ideally we would now match our activity-sensitive lines to specific atomic transitions, but this is difficult as there really is no such thing as a spectral line in isolation. Many of our activity-sensitive lines are blends with weaker spectral lines, and a few are blends between two lines of similar strength.
To get line species and solar line depths, we used a Vienna Atomic Line Database (VALD) \citep{piskunov95} ``extract stellar" query where we input $\alpha$~Cen~B atmospheric parameters\footnote{The ``extract stellar" star parameters we used were: microturbulence: 1.1~km~s$^{-1}$, $T_{\rm eff}$: 5200~K, $\log{g}$: 4.5~(g in cgs units), and chemical composition: ``Fe: -4.34". The last parameter is VALD's default stellar iron abundance plus 0.2 dex \citep{portodemello08}. VALD did not have an exact stellar model for $\alpha$~Cen~B, so it based the query output on model castelli\_ap00k2\_T05250G45.krz \citep{castelli04}.} \citep{portodemello08} and a line detection threshold of 0.1. Since we cannot be sure which line in a given wavelength interval is responding to activity, we consider all VALD lines within $2 \times 10^{-5} \lambda$ ($\sim 0.1 \AA$) of line center as candidates for the activity signal.
Table~\ref{lines} shows each line's core flux correlation coefficient, wavelength, slope of best-fit line to core flux vs S-index plot for each star, VALD atomic species and corresponding solar line depths, and previous literature discussing the line.

64 of the 154 lines with line core flux correlations $| \tau | < 0.5$ had coefficients $| \tau | \geq 0.3$, so we suspect these are activity-sensitive lines with noisier core fluxes, making them less useful for diagnosing activity. For reference, we include two of our four lines that had half-depth range correlation coefficients $| \tau | \geq 0.5$ in Table~\ref{lines}, Fe $5250.2 \AA$ and Na $5890.0 \AA$, as these have moderate core flux correlation coefficients as well as literature references where they are used to diagnose activity. While our center of mass measurements did not yield any lines with correlation coefficients $| \tau | \geq 0.5$, they did find 16 lines with $| \tau | \geq 0.3$, 6 of which had $| \tau | \geq 0.4$. Interestingly, exactly half (8/16 and 3/6) of these had positive correlation coefficients, so there does not appear to be a preference for redshifts or blueshifts in line center of mass, whereas activity-sensitive lines identified by the other two metrics almost always had positive correlation coefficients.

Next we assess line list completeness, demonstrate that we can recover the star's rotation period from the line core fluxes, and compare each line's behavior in the spectra of $\alpha$~Cen~B vs.\ $\epsilon$~Eri.

%\clearpage

%\scriptsize
%\onecolumn
%\small
%\begin{longtable}{lllllp{0.25\linewidth}}
\begin{table}
%\caption*{Activity-Sensitive Lines\label{lines}}\\
\caption{Activity-Sensitive Lines\label{lines}}
\footnotesize
\begin{tabular}{llllll}
%\hline\hline
\tableline\tableline
$\tau$ & Wavelength & $\epsilon$~Eri Slope\tablenotemark{a} & $\alpha$~Cen~B Slope\tablenotemark{a} & Species (VALD depth) & References\\
%\hline
\tableline
0.753&5110.42&0.212&0.137&Fe I (0.93), Fe I (0.72), Fe I (0.21)&\\
0.733&4375.94&0.205&0.14&Fe I (0.97), Fe I (0.8), Ce II (0.14)&\\
0.71&4427.32&0.185&0.13&Fe I (0.97), Fe I (0.82), V I (0.11)&\\
0.702&4461.66&0.176&0.11&Fe I (0.96)&\\
0.694&5012.08&0.155&0.092&Fe I (0.93), Fe I (0.67)&\\
0.685&5269.54&0.102&0.11&Fe I (0.94)&\\
0.676&5397.13&0.142&0.118&Fe I (0.93), Fe I (0.52), Ti I (0.34)&\\
0.673&5429.7&0.129&0.132&Fe I (0.92)&\\
0.667&5506.78&0.132&0.114&Fe I (0.9)&\\
0.667&4571.1&0.147&0.073&Mg I (0.92), Cr I (0.14)&4, 5, 8, 11\\
0.661&6562.81&0.149&0.444&H I (0.45) (H $\alpha$)&1, 2, 3, 6, 10, 11\\
0.645&5051.64&0.134&0.092&Fe I (0.92)&\\
0.643&5371.5&0.099&0.12&Fe I (0.93), Fe I (0.54), Cr I (0.17)&\\
0.639&5405.78&0.105&0.121&Fe I (0.93)&\\
0.634&5501.47&0.127&0.082&Fe I (0.89)&\\
0.631&5227.19&0.097&0.089&Fe I (0.93), Fe I (0.88)&\\
0.625&5107.45&0.133&0.08&Fe I (0.91), Al I (0.11)&\\
0.616&5434.53&0.117&0.118&Fe I (0.92)&\\
0.613&5432.54&0.182&0.177&Mn I (0.74)&\\
0.586&5497.52&0.116&0.097&Fe I (0.9)&\\
0.583&5194.95&0.118&0.108&Fe I (0.91)&\\
0.576&5895.93&0.062&0.05&Na I (0.9)&7, 10, 11\\
0.544&4827.46&0.212&0.126&V I (0.63)&\\
0.54&5083.34&0.115&0.054&Fe I (0.92)&\\
0.527&6013.49&0.138&0.135&Mn I (0.59)&\\
0.525&6252.56&0.106&0.084&Fe I (0.8)&\\
0.522&5167.33&0.082&0.089&Mg I (0.92)&8, 11\\
0.516&5150.85&0.133&0.096&Fe I (0.91), Mn I (0.44)&\\
0.515&4994.14&0.116&0.088&Fe I (0.92)&\\
0.514&6191.57&0.098&0.084&Fe I (0.82)&\\
0.513&6393.61&0.108&0.089&Fe I (0.81)&\\
0.51&6430.85&0.105&0.063&Fe I (0.79)&\\
-0.509&4861.33&-0.169&0.113&H I (0.49) (H $\beta$)&2, 10, 11\\
0.507&5420.35&0.118&0.134&Mn I (0.65)&\\
0.503&4426.02&0.173&0.091&V I (0.75), Ti I (0.67)&\\
0.502&5707.0&0.142&0.07&V I (0.5), Fe I (0.42)&\\
0.5&5172.69&0.059&0.059&Mg I (0.93)&5, 8, 11\\
0.5&4602.95&0.086&0.056&Fe I (0.94)&\\
0.5&6230.73&0.088&0.053&Fe I (0.82), V I (0.46)&\\
0.492\tablenotemark{b}&5183.61&0.051&0.042&Mg I (0.94), Ti II (0.11)&8\\
0.472\tablenotemark{c}&5250.21&0.127&0.153&Fe I (0.85)&9\\
0.416\tablenotemark{b,c}&5889.96&0.033&0.043&Na I (0.9)&7, 10, 11\\
%\hline
\tableline
%& \\
\end{tabular}
\tablenotetext{a}{Slope of line fitted to core flux vs.\ S-index plot (e.g. Figure~\ref{fig5}) for $\epsilon$~Eri or $\alpha$~Cen~B.}
\tablenotetext{b}{Member of the Na doublet or Mg triplet, both popular activity indicators in the literature (see references).}
\tablenotetext{c}{Line Core flux correlation coefficient is less than 0.5, but half-depth range correlation coefficient is greater than 0.5.}
\tablerefs{(1)~\citealt{barnes14}; (2)~\citealt{cram79}; (3)~\citealt{cram85}; (4)~\citealt{langangen09}; (5)~\citealt{mauas88}; (6)~\citealt{robertson13}; (7)~\citealt{robertson15}; (8)~\citealt{sasso17}; (9)~\citealt{stenflo73}; (10)~\citealt{thatcher91}; (11)~\citealt{thatcher93}.}
\end{table}
%\clearpage
%\twocolumn
\normalsize
%\clearpage

\section{Completeness, periodicity, and behavior in $\alpha$~Cen~B vs.\ $\epsilon$~Eri}\label{additional}

To determine the completeness of our activity-sensitive line list, we first computed a master spectrum for $\epsilon$~Eri by taking the median of all 540 spectra after RV shifting, linearly interpolating them onto a common wavelength grid and continuum-normalizing them (as described in Appendix~\ref{norm}). We then produced the median normalized 1-D spectrum by combining the overlapping edges of adjacent orders with a weighted average. To get the weights, first we divided each RV shifted, interpolated and non-normalized spectrum by its flattened trimmed mean flux (10th to 90th percentile) to roughly match the continua of these non-normalized spectra. At each wavelength, we used the median flux of these continuum matched spectra, which are a central estimate of the relative flux a given data point receives, as the weight for our averaging of adjacent order edges. To produce the 1-D wavelength array, wavelength entries are copied from the appropriate order, with overlapping wavelength regions split into two equal-length halves. For example, a $10 \AA$ wide overlapping region would draw values within the smaller $5 \AA$ range from the shorter-wavelength spectral order, and values within the larger $5 \AA$ range from the longer-wavelength spectral order.

\begin{figure}
\epsscale{1.0}
\centering
\plotone{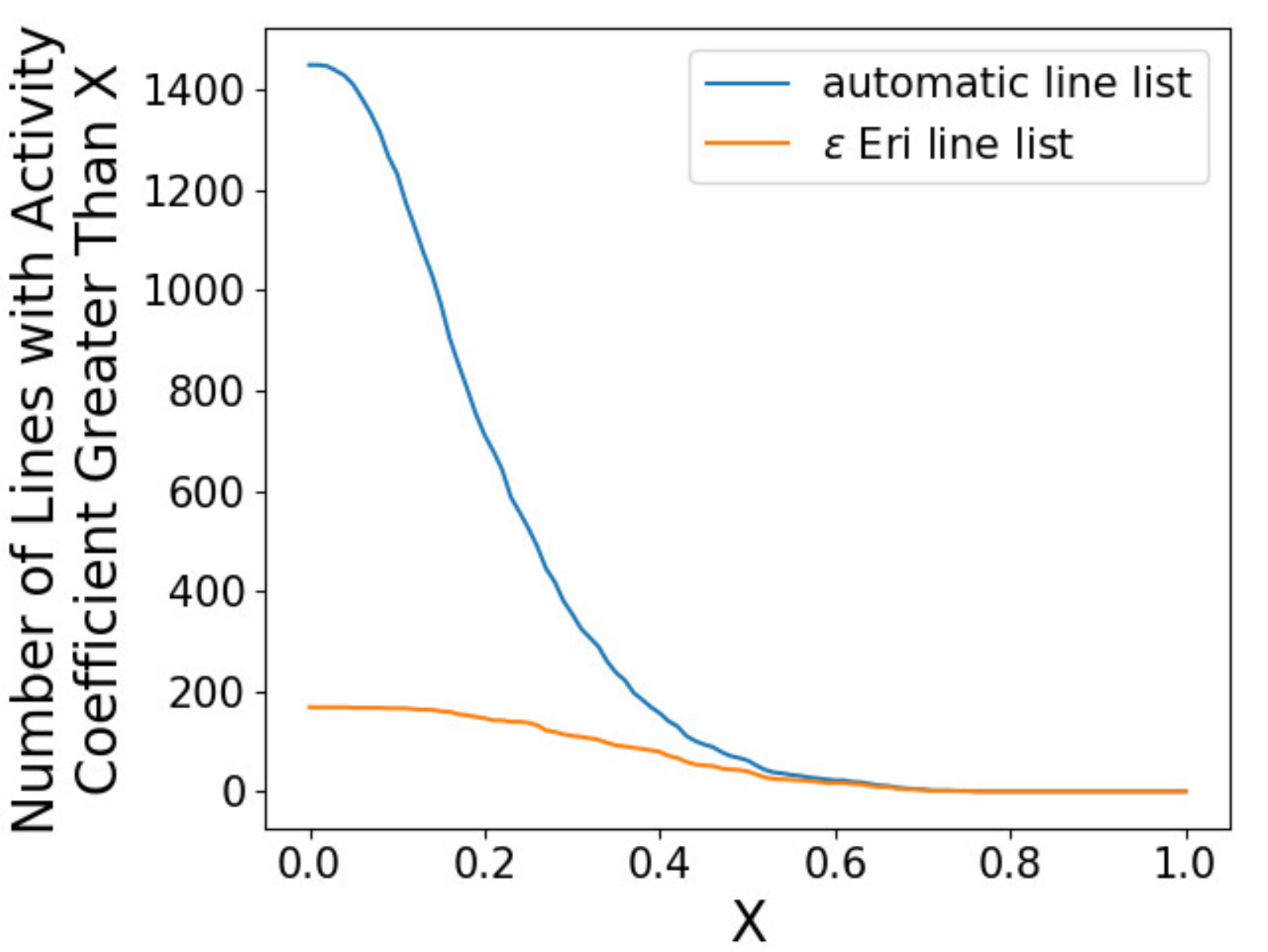}
\caption{Plot showing completeness of our visual identification of activity-sensitive lines in Section~\ref{EpsEri}. The two curves are very close together above an activity coefficient of X=0.5, suggesting our visual identification of activity-sensitive lines is complete down to X=0.5.\label{fig6}}
\end{figure}

Using our normalized median 1-D spectrum of $\epsilon$~Eri, a line list was generated by picking all pixels in the spectrum below a value of 0.8, grouping all such selected pixels within $\Delta \lambda / \lambda = 10^{-5}$ of their nearest selected neighbor into ``lines", and then rejecting any line whose minimum flux was located at its minimum or maximum wavelength, or whose minimum flux was greater than $\sim$0.6\footnote{To be more precise, the line's half-depth flux was required be less than the flux of the pixels at the line edges. In the limit of very small pixel size, both pixels at the line edges have a flux of 0.8, so the line flux minimum was required to be less than or equal to 0.6. This makes sure the pixels in each line have a range of at least 0.2 in normalized flux.} (had a line depth $<40$\%). Figure~\ref{fig3} shows three such automatically identified lines in blue. We then computed the line core flux, half-depth range, and center of mass, along with each metric's correlation coefficient with the S index, for the $\sim 1500$ lines detected by our grouping algorithm in the $\alpha$~Cen~B spectra. For each spectral line, we chose the absolute value of the highest of the three correlation coefficients, which we labeled ``activity coefficient'' X. In Figure~\ref{fig6}, we plot the number of lines with an activity coefficient greater than a given cutoff, versus that cutoff, for both the automatically generated line list and the lines of interest from Section~\ref{EpsEri}. The merger of the two curves around a cutoff of 0.5 suggests that our visual identification of lines of interest is complete for correlation coefficients $| \tau | \geq 0.5$.

It is worth noting that nearly all lines in the spectrum have a correlation coefficient of at least 0.1. This could result from continuum emission filling in the line cores, which would affect all of the lines in the spectrum, especially deeper lines. However, we looked at a plot of average line core flux vs. line core flux correlation coefficient (not shown in this paper) and did not see any relationship, so we suspect this continuum filling signal is small (less than 10\%) compared to chromospheric emission or changing absorption probabilities. Since these line core fluxes are changing independently of a continuum emission source, our half-depth range metric is more orthogonal to our line core flux metric than line FWHM (full width at half maximum) would be, as FWHM varies with line core flux since it depends on where the half-maximum (and hence the maximum depth) of the line is, but our half-depth range uses a single half-depth flux measured from the median of all our $\epsilon$~Eri spectra.

\begin{figure}
\centering
\includegraphics[scale=0.45]{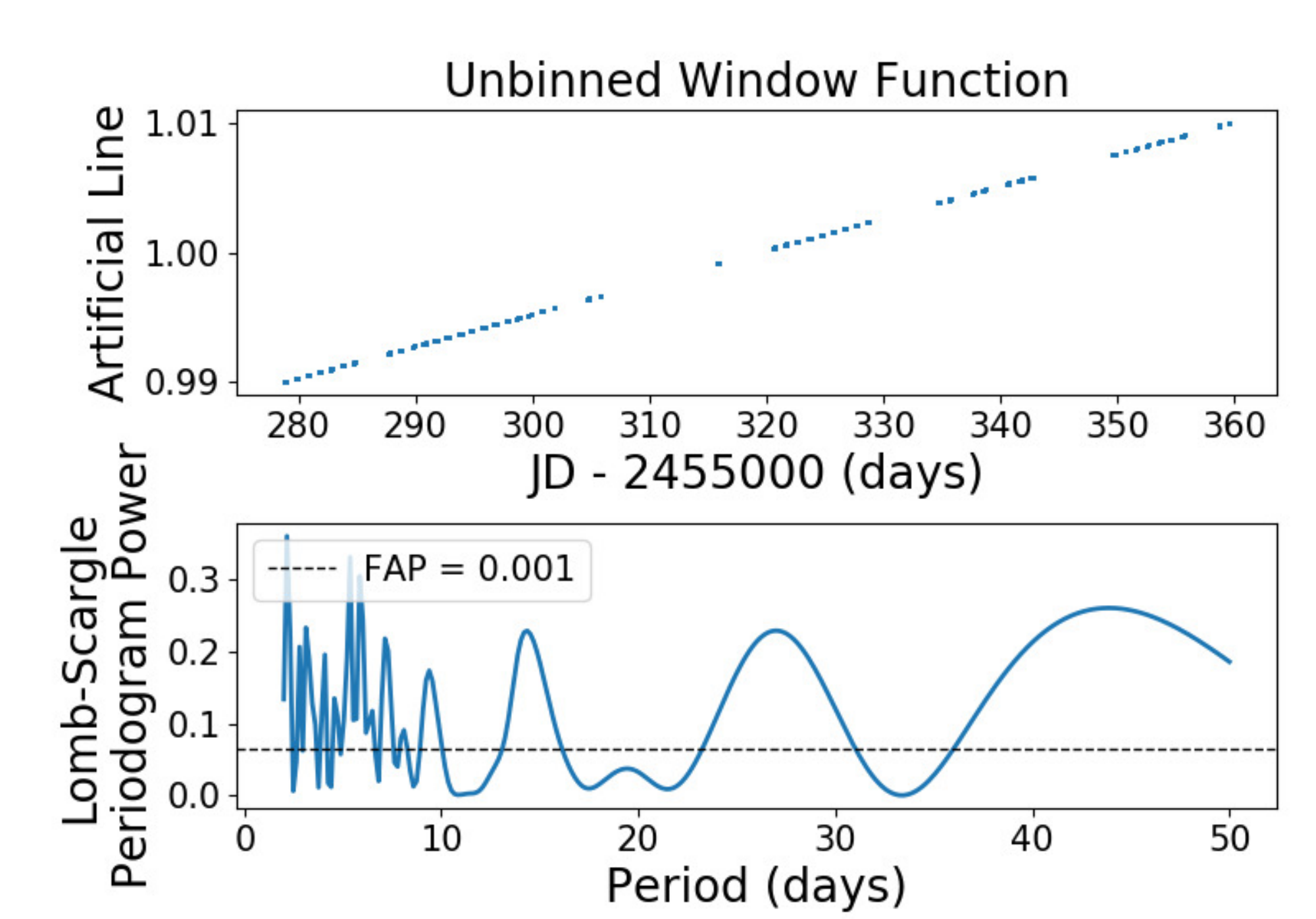}
\includegraphics[scale=0.45]{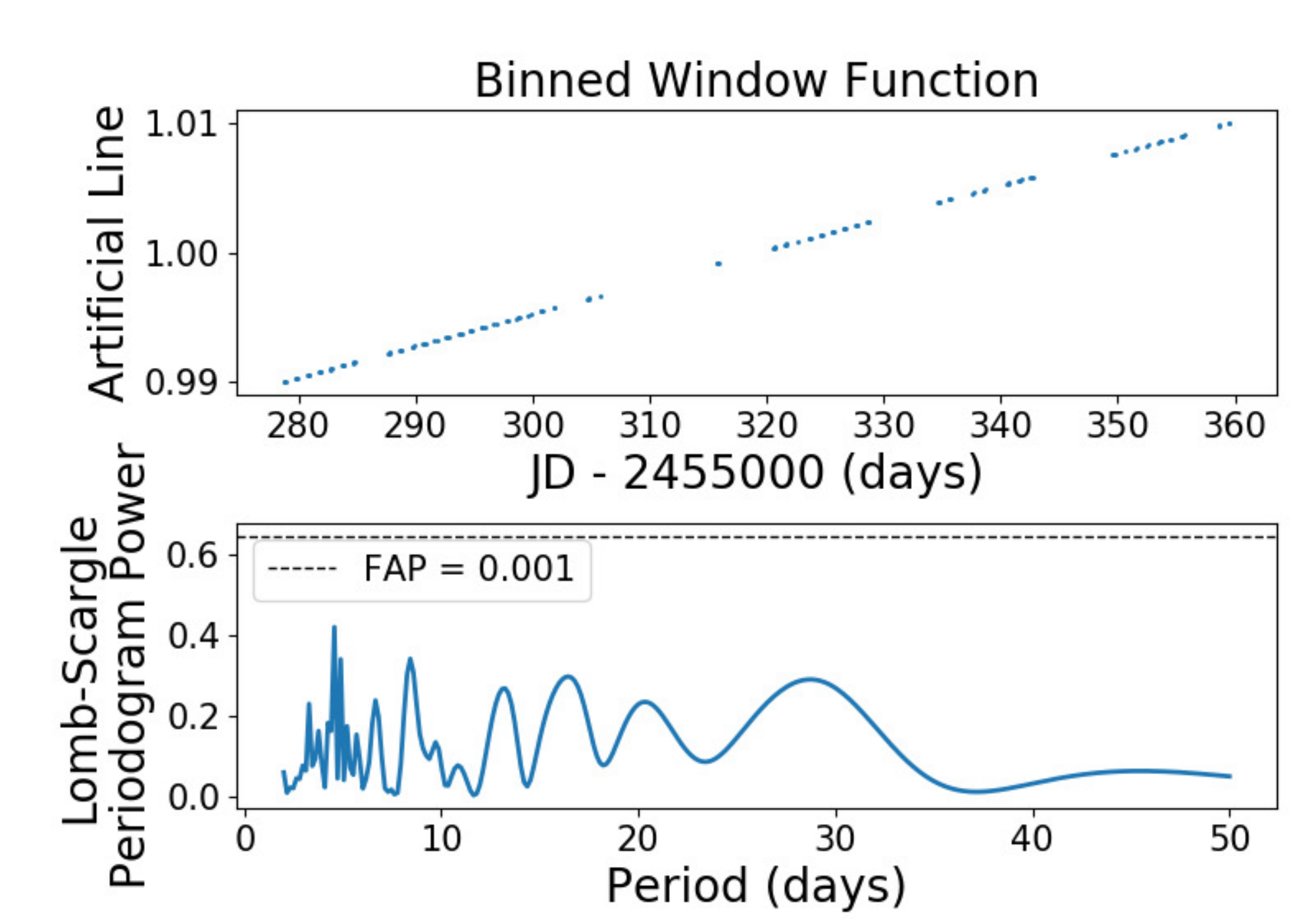}
\caption{Time series and generalized Lomb-Scargle Periodograms for the window functions of our unbinned measurement times (top) and binned measurement times (bottom). The unbinned window function contains many significant peaks, which could lead us to overestimate the significance of periods found in our data. However, the binned window function no longer has any peaks above the p=0.001 significance level, so we can be more confident in our signficance estimates for period detections. \label{fig7}}
\end{figure}

\begin{figure}
\centering
\includegraphics[scale=0.45]{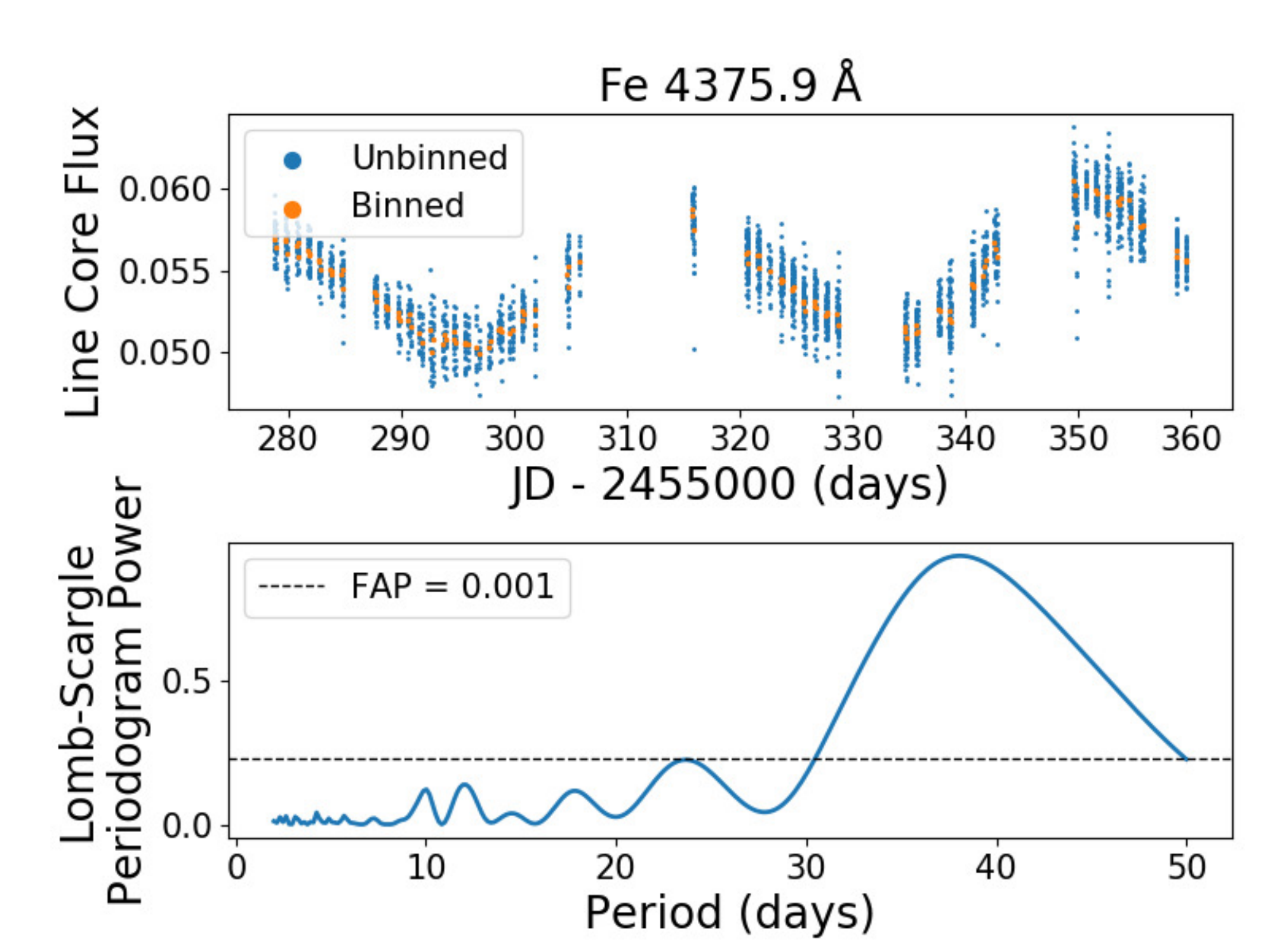}
\includegraphics[scale=0.45]{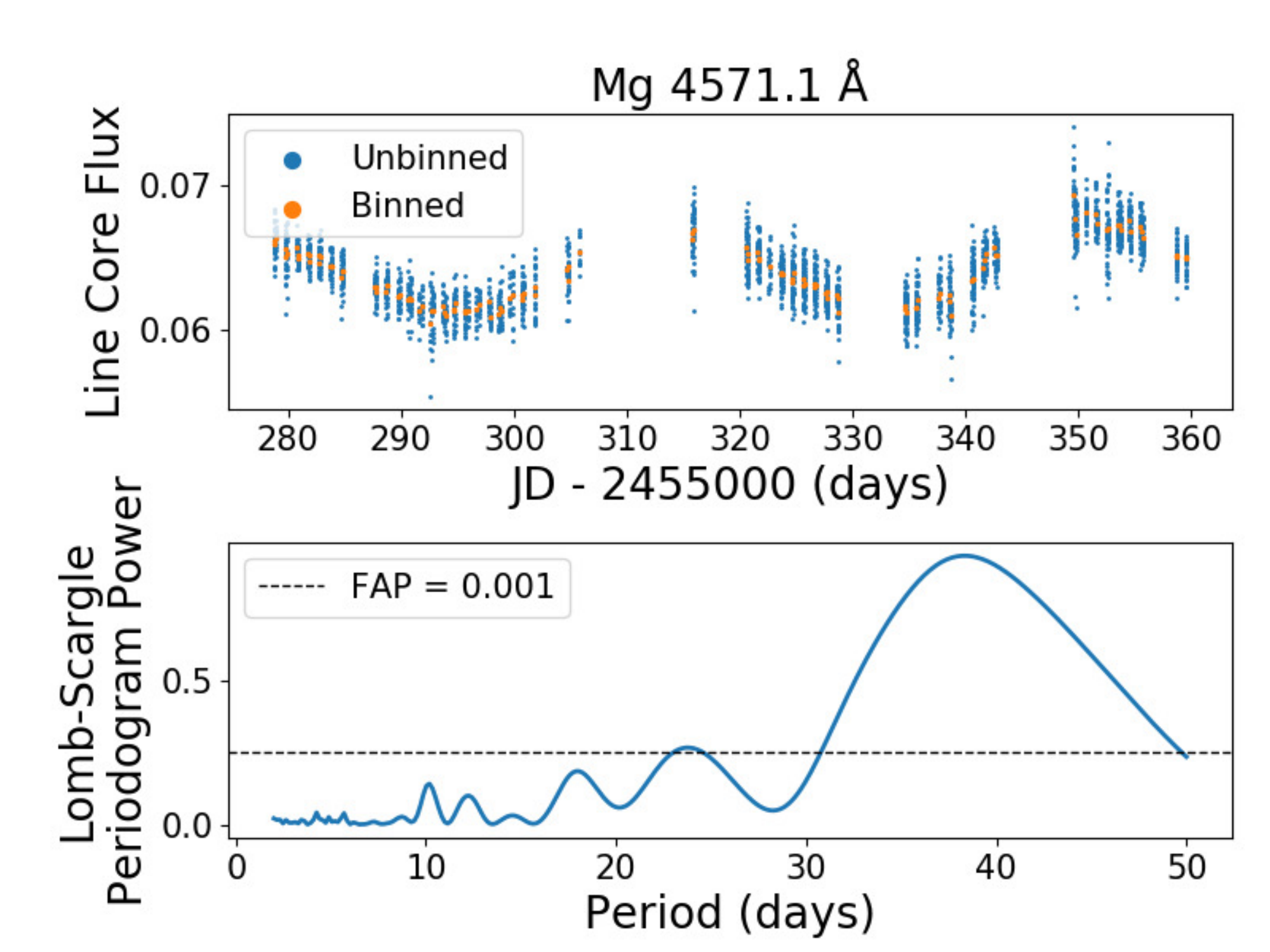}
\includegraphics[scale=0.45]{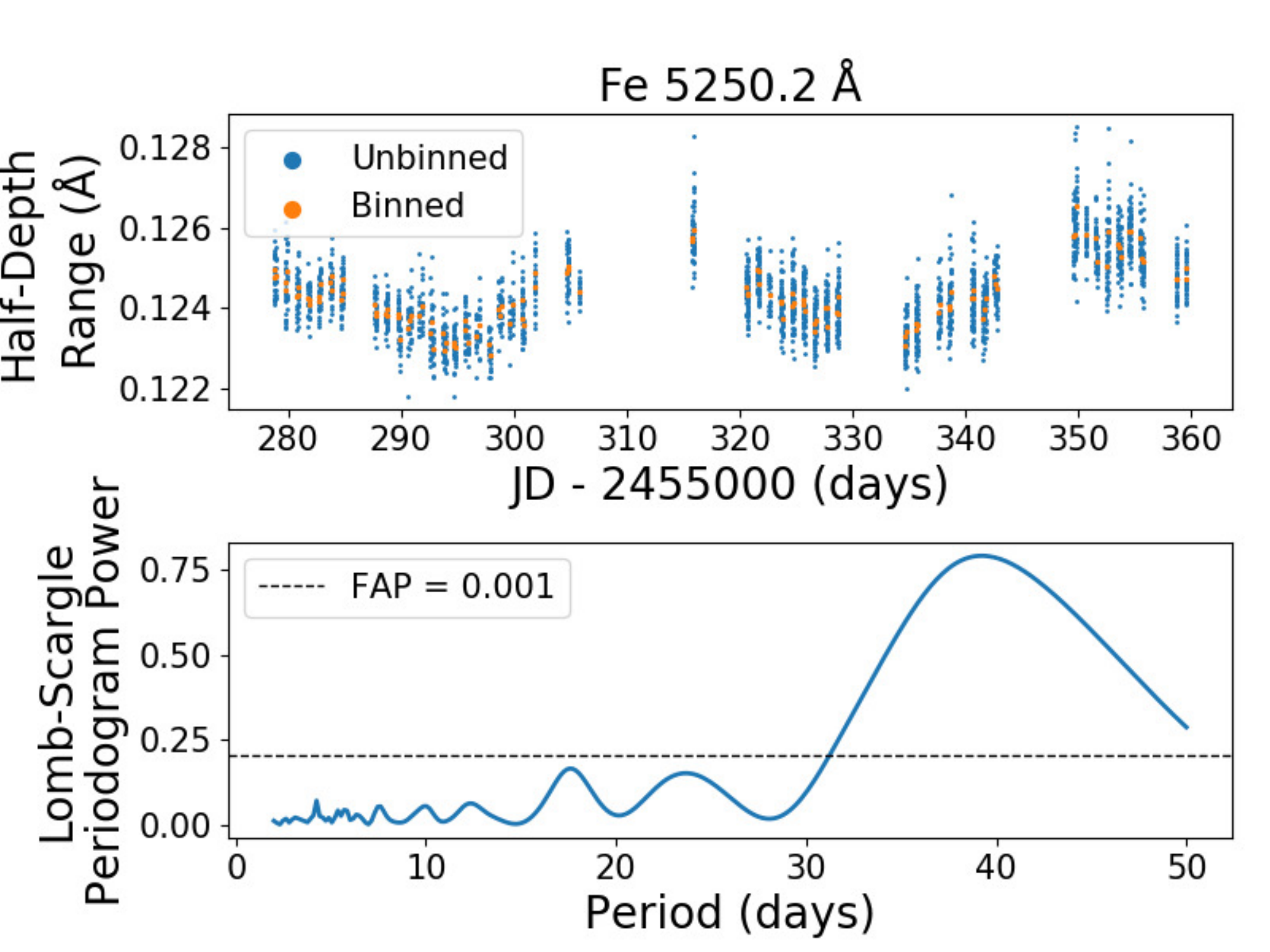}
\caption{Time series and generalized Lomb-Scargle Periodograms of selected line properties for lines at $4375.9 \AA$, $4571.1 \AA$, and $5250.2 \AA$. \label{fig8}}
\end{figure}

To confirm that our comparisons with S-index are indeed finding periodic variability in our activity-sensitive lines, we computed generalized Lomb-Scargle periodograms\footnote{using the python class \texttt{astropy.stats.LombScargle}, for details see http://docs.astropy.org/en/stable/stats/lombscargle.html} \citep{zechmeister09,pricewhelan18}. Initially we computed periodograms using the unbinned line property measurements shown in our correlation plots in Figure~\ref{fig5}. However, these periodograms contained many significant peaks that were due to the window function of the unbinned data (FIgure~\ref{fig7}, top). We call a periodogram peak significant when it has estimated false alarm probability (FAP) $p < 0.001$. To remove periodogram peaks caused by our window function, we binned each time series of 2491 line property measurements using a one hour sliding window into 133 mean line property measurements. We also calculated the standard deviation within each bin to estimate the error in these mean line property measurements. The binned window function no longer contains any significant peaks (Figure~\ref{fig7}, bottom), so we can be confident that peaks we detect in our binned line property measurements are really due to periodic modulation of line properties. In each periodogram, we calculate the periodogram power corresponding to $p = 0.001$ using the bootstrap method. However, the bootstrap method requires computing approximately $10/p$ individual periodograms, making it computationally expensive for very low FAPs. Hence we estimated the FAP of each of our significant ($p < 0.001$) periodogram peaks using the algorithm by \cite{baluev08}.
%We detrended by dividing each line property time series by a quadratic polynomial fit, then multiplying by the time series' mean to recover the variability amplitude. After experimenting with linear, quadratic and cubic polynomial detrending, we chose quadratic because it can capture the long-term magnetic variability of $\alpha$~Cen~B \citep{dumusque12} better than linear detrending, without over-fitting and significantly reducing the amplitude of line core flux variations as in cubic polynomial detrending.
Table~\ref{periods} lists the period as measured by the line core flux and corresponding FAP for each activity-sensitive line, and Figure~\ref{fig8} shows time series and generalized Lomb-Scargle periodograms for the same three activity-sensitive lines as shown in the correlation plots in Figure~\ref{fig5}. All periodogram peaks are between 35 and 40 days, which agrees well with the star's $36.2 \pm 1.4$ day rotation period published by \cite{dewarf10}. Note the estimated FAPs are all below $10^{-25}$, so we have good reason to believe all of these signals are truly periodic.

%Given that $\alpha$~Cen~B's rotation period is $\sim 36$ days \citep{dewarf10,dumusque12}, our measured line core fluxes appear to trace some combination of rotation and spot configuration changes.

\begin{table}
\caption{Line Core Flux Periods and FAPs for Activiy-Sensitive Lines\label{periods}}
\footnotesize
\begin{tabular}{lllllllll}
\tableline
\tableline
Wavelength & Period & FAP & Wavelength & Period & FAP & Wavelength & Period & FAP\\
\tableline
4375.94&38.09&6.4e-74&5150.85&37.46&1.5e-56&5434.53&38.35&3.2e-62\\
4426.02&36.25&1.4e-34&5167.33&38.23&2.6e-49&5497.52&38.16&1.4e-57\\
4427.32&37.72&4.1e-74&5172.69&38.03&6.0e-45&5501.47&37.90&4.4e-59\\
4461.66&37.91&7.1e-73&5183.61&38.65&5.7e-38&5506.78&37.85&5.4e-80\\
4571.10&38.33&3.0e-73&5194.95&37.96&2.4e-41&5707.00&37.15&6.5e-49\\
4602.95&37.55&1.2e-57&5227.19&38.30&1.1e-57&5889.96&35.25&1.3e-30\\
4827.46&37.82&3.1e-50&5250.21&38.12&9.5e-39&5895.93&39.83&2.3e-44\\
4861.33&39.76&7.2e-26&5269.54&38.54&5.1e-72&6013.49&38.06&2.0e-43\\
4994.14&38.20&1.5e-67&5371.50&38.10&7.4e-69&6191.57&38.49&4.4e-49\\
5012.08&38.18&3.9e-66&5397.13&38.32&1.2e-69&6230.73&37.51&1.7e-46\\
5051.64&38.09&7.7e-66&5405.78&38.26&3.4e-73&6252.56&37.84&3.7e-45\\
5083.34&38.56&3.4e-54&5420.35&37.16&3.5e-41&6393.61&37.94&1.2e-51\\
5107.45&38.36&9.4e-64&5429.70&38.12&1.1e-71&6430.85&38.16&5.3e-45\\
5110.42&38.17&5.2e-73&5432.54&37.12&7.8e-49&6562.81&38.14&1.2e-49\\
\tableline
\end{tabular}
\end{table}
\normalsize

\begin{figure}
\epsscale{1.0}
\centering
\plotone{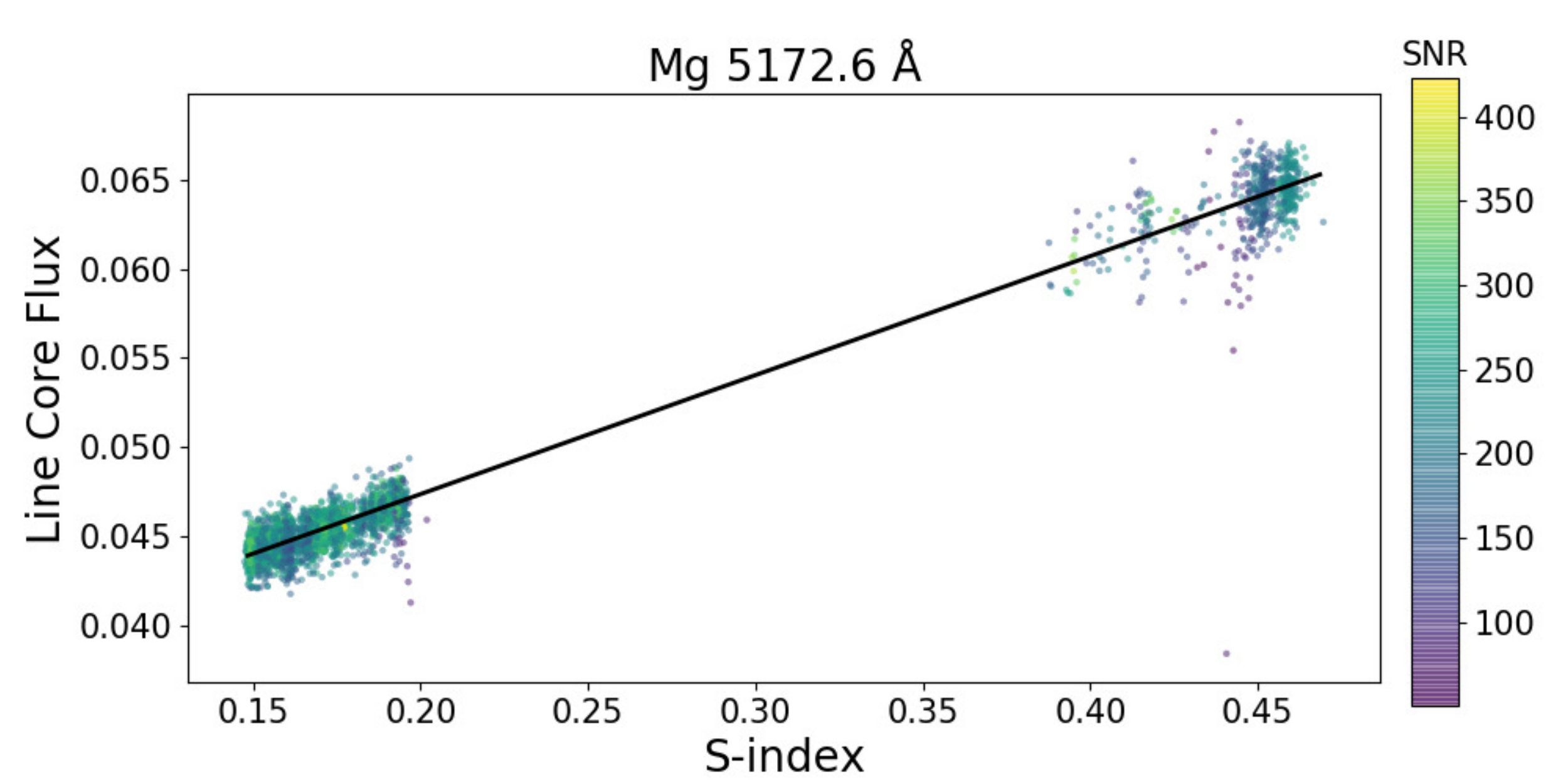}
\caption{Combined $\epsilon$~Eri + $\alpha$~Cen~B correlation plot for Mg 5173. The same line core flux vs. S-index slope describes the line's behavior in both stars. Points are color-coded by signal-to-noise ratio in the line core flux.\label{fig9}}
\end{figure}

From Table~\ref{lines}, we notice that a few of our activity-sensitive lines, such as the Mg line at $5173 \AA$, have correlation plot slopes for $\epsilon$~Eri and $\alpha$~Cen~B that are similar. This prompted us to plot the line core flux vs.\ S-index for both stars on a single correlation plot, as shown in Figure~\ref{fig9}. Interestingly, a single line fits Mg 5173's line core fluxes for both stars quite well, without any shifting in the y-direction, suggesting the shallower line depth (consistent with infilling from chromospheric emission or changing absorption probability) in $\epsilon$~Eri may be solely due to activity, not any underlying differences in abundances. However, most other spectral lines require a break in the core flux vs.\ S-index slope to fit the data from both stars, indicating that stellar activity response is generally more complex than what a linear fit can capture.

\section{Discussion and conclusions}\label{discussion}

We searched archival HARPS spectra of $\epsilon$~Eri and $\alpha$~Cen~B for optical spectral lines whose variation correlates with the Mt.\ Wilson S-index, a measure of chromospheric activity. After we visually identified lines of interest based on $|({\rm A-Q})|$ correlating with S-index, we measured three properties for each line of interest: core flux, half-depth range, and center of mass (see \S~\ref{AlphaCenB}). We found that core flux was the most robust quantitative indicator of stellar activity, as measured by the Kendall's $\tau$ correlation coefficient with S-index, for our visually identified lines of interest. In addition to our methods for identifying activity-sensitive lines, we also present a list of 39 lines whose core fluxes strongly correlate with S-index (see Table~\ref{lines}). We also note that Fe~$5250 \AA$, which Zeeman splits in sunspots, shows a strong correlation between half-depth range and S-index.

Our list of activity-sensitive lines may be used to improve methods of disentangling the RV signals of exoplanets from RV signals of stellar activity. \cite{giguere16} note that their HH' method, an H$\alpha$-based variation on the FF' method \citep{aigrain12} for removing RV variations due to stellar activity from planet-search data, could be improved by incorporating more activity-sensitive lines. Similarly, the multiple activity indicators used to gauge stellar activity in \cite{rajpaul15} could be expanded to include activity-sensitive line depths. However, these methods of trying to separate components of RV time series will continue to perform poorly when an exoplanet period is similar to the stellar activity signal period. One potential solution is directly calculating convective blueshift using stellar effective temperatures and activity levels \citep{meunier17}. Another possibility is using photospheric activity indicators, not just chromospheric activity indicators, to improve the RV measurement process \citep{davis17}. Half-depth range, which we believe probes Zeeman splitting in some lines, may be a promising photospheric activity metric.

Although the activity-sensitive lines we have identified are at optical wavelengths ($4376 \AA$ to H$\alpha$), the methods we used to find them are not particular to only one wavelength range. A similar procedure that unites visual inspection with automated line property measurements could be used to find stellar activity-sensitive lines in any wavelength range that contains a previously known stellar activity indicator.

\acknowledgments

The authors thank Debra Fischer for ideas that inspired this work. The authors also thank Lars Buchhave, Heather Celga, Xavier Dumusque, and Jeff Valenti for ideas that helped make this work possible.
This research is based on data products from observations made with ESO Telescopes at the La Silla Paranal Observatory under programme IDs 60.A-9036(A), 072.C-0488(E), 072.C-0513(B), 072.C-0513(D), 074.C-0012(A), 074.C-0012(B), 076.C-0878(A), 077.C-0530(A), 078.C-0833(A), 079.C-0681(A), 084.C-0229(A), 085.C-0318(A), and 192.C-0852(A).
This research has made use of the services of the ESO Science Archive Facility, and is based on data obtained from the ESO Science Archive Facility under request numbers 243402, 243835, 300034, 300050. This work has made use of the VALD database, operated at Uppsala University, the Institute of Astronomy RAS in Moscow, and the University of Vienna. This research has made use of the SIMBAD database, operated at CDS, Strasbourg, France. We acknowledge support from the University of Delaware Department of Physics and Astronomy for providing computing resources through the Farber and Mills computing clusters at University of Delaware. We also acknowledge support from NSF CAREER award 1520101, and the Delaware Space Grant College and Fellowship Program, NASA Grant NNX15AI19H.

\clearpage

\appendix

\section{Automated Continuum Normalization}\label{norm}

In Section~\ref{AlphaCenB} we mentioned that all $\alpha$~Cen~B spectra in our sample are continuum normalized. To accomplish this, we developed an iterative mask estimation (IME) algorithm to mask out spectral lines while fitting a continuum function. Our IME algorithm estimates a regression function (here a spectral order's continuum) in the presence of heteroscedastic noise (here spectral lines) by iteratively masking outliers (here flux values in spectral lines) based on coefficients of determination from sets of leave-one-out regressions (here least squares). Below we describe our implementation of IME to normalize all of the HARPS spectra mentioned in this paper.

\subsection{Initialization}

Each HARPS spectrum has 72 orders and each order is 4096 pixels across. The original echelles have already been reduced to 72x4096 arrays when we retrieve them from the ESO archive (we use the e2ds files). Wavelength calibrations and blaze functions associated with each spectrum are also obtained from the ESO archive. To initialize each automated continuum normalization, each spectrum is paired with its associated wavelength calibration and divided by its associated blaze function. All remaining steps are carried out on the 4096 pixels in each order of each spectrum independently of any other orders or spectra.

First, to speed up the continuum fits, we reduce our IME input data from 4096 pixels to 100 ``local maximum pixels" (LMPs). The 4096 pixels in each blaze-corrected order are divided into 100 consecutive subsets: the first 99 subsets each have 41 pixels, and the last subset has 37 pixels. For each of the 100 subsets of pixels, we find the pixel of maximum flux and record its wavelength and flux ($\lambda$, $F_0$) before division by the blaze function, and flux after division by the blaze function ($F_B$). We use subsets of these LMPs in the next step (\S~\ref{LMPs}) to fit the continuum with 2nd order polynomials using weighted least-squares regressions.

We found a 2nd order polynomial to be optimal after dividing by the HARPS blaze functions provided on the ESO archive. We recommend a 4th to 6th order polynomial for those who choose not to use the ESO-provided blaze functions. The 2nd order polynomial least squares regressions use the non blaze-corrected flux ($F_0$) for each LMP as a weighting factor. This ensures that pixels with fewer counts toward the ends of orders, where the signal-to-noise ratio (SNR) is lower, contribute less to the continuum fits. 

\subsection{Selecting LMPs for Fitting}\label{LMPs}

Our polynomial fits use a mask which lists the LMPs that should be ignored when fitting. LMPs are usually added to the mask because they lie on spectral lines; we describe in detail how members of the mask are chosen below. LMPs are added to the mask iteratively, reducing the total number of points being fit, until the weighted $R^2$ (coefficient of determination) satisfies $|1-R^2| < t$, where $t$ is the tolerance. We initialize the tolerance to $t=0.005$ for most HARPS orders (and we will describe how we found custom tolerances for the bluest 28 orders in Section~\ref{tolerance}).

We begin by using the \texttt{scipy.optimize.curve\_fit} function to fit a continuum polynomial to all LMPs using least squares. Next, we search for mask candidates that satisfy all of the following criteria, which force the fit to converge toward the continuum:
\begin{enumerate}
\item LMP not currently in mask.
\item LMP flux is less than the polynomial value at the LMP wavelength, or CRA (cosmic ray allowance) $<$ 2.
\item The order side the LMP is on (left or right) contains at least 5 unmasked LMPs.
\end{enumerate}

Criteria 2 and 3 are specifically tailored to the spectral orders to which we are fitting continuum polynomials to, and would not be included in more general uses of IME.  CRA starts at 0, and gets incremented by 1 every time a new LMP with flux greater than the current polynomial value is masked out. That way, our polynomial fitting algorithm masks out no more than 2 LMPs with fluxes above the current fitting function. Most of the masked LMPs will come from regions of low flux, which could be parts of broad absorption lines and should not be used in continuum fits. Note that LMPs with flux greater than the polynomial fit are not necessarily affected by cosmic rays, but masking out only 2 out of 100 LMPs from above the fitting function still allows for the vast majority of points in the fit to be forced upward towards continuum. Each order is broken into two sides: left (LMP indices 0-49) and right (50-99). Having a minimum number of unmasked LMPs in each side of the order ensures the fitting polynomials cover a large wavelength range in each order, while allowing the flexibility to mask out complex absorption line patterns.

LMPs are added to the mask by repeating the following leave-one-out procedure on each mask candidate:
\begin{enumerate}
\item Candidate LMP is added to the mask (temporarily)
\item The polynomial fit is performed on all unmasked LMPs, and the $R^2$ for that fit is calculated
\item Mask candidate is removed from the mask
\end{enumerate}

After $R^2$ is calculated for each mask candidate, the candidate which results in the minimum $R^2$ is accepted, and joins the mask permanently. This procedure is repeated until $|1-R^2| < t$.

\subsection{Assigning a Tolerance to Each Order}\label{tolerance}

Since line blanketing and signal-to-noise ratio vary between spectral orders, it is not always possible to find a polynomial fit to the continuum with $t \leq 0.005$. Here we describe our procedure for finding an appropriate tolerance for the bluest 28 orders of the HARPS spectra. We perform this procedure using a single spectrum with no visually identifiable irregularities, and apply the results to all 3031 HARPS spectra in the $\epsilon$~Eri and $\alpha$~Cen~B samples.

\begin{enumerate}
\item A tolerance of 0.005 is assigned to each order.
\item The above procedure in \S~\ref{LMPs} for fitting the continuum of each spectral order is carried out, but without the minimum of 4 unmasked LMPs per side.
\item If the resulting mask is removing more than 80\% of the maximum number of possible LMPs, which is 100 minus the number of coefficients in the polynomial, the tolerance for that spectral order is doubled.
\item Steps 2 and 3 are repeated for spectral orders whose tolerance was doubled, using the updated tolerances on each new iteration until the condition is step 3 is no longer met for any order.
\end{enumerate}

In our implementation of this procedure on HARPS spectra, orders redder than number 28 did not have their tolerances increased beyond 0.005 by the above steps.

\subsection{Concluding remarks}

\begin{figure}
\epsscale{1.0}
\centering
\plotone{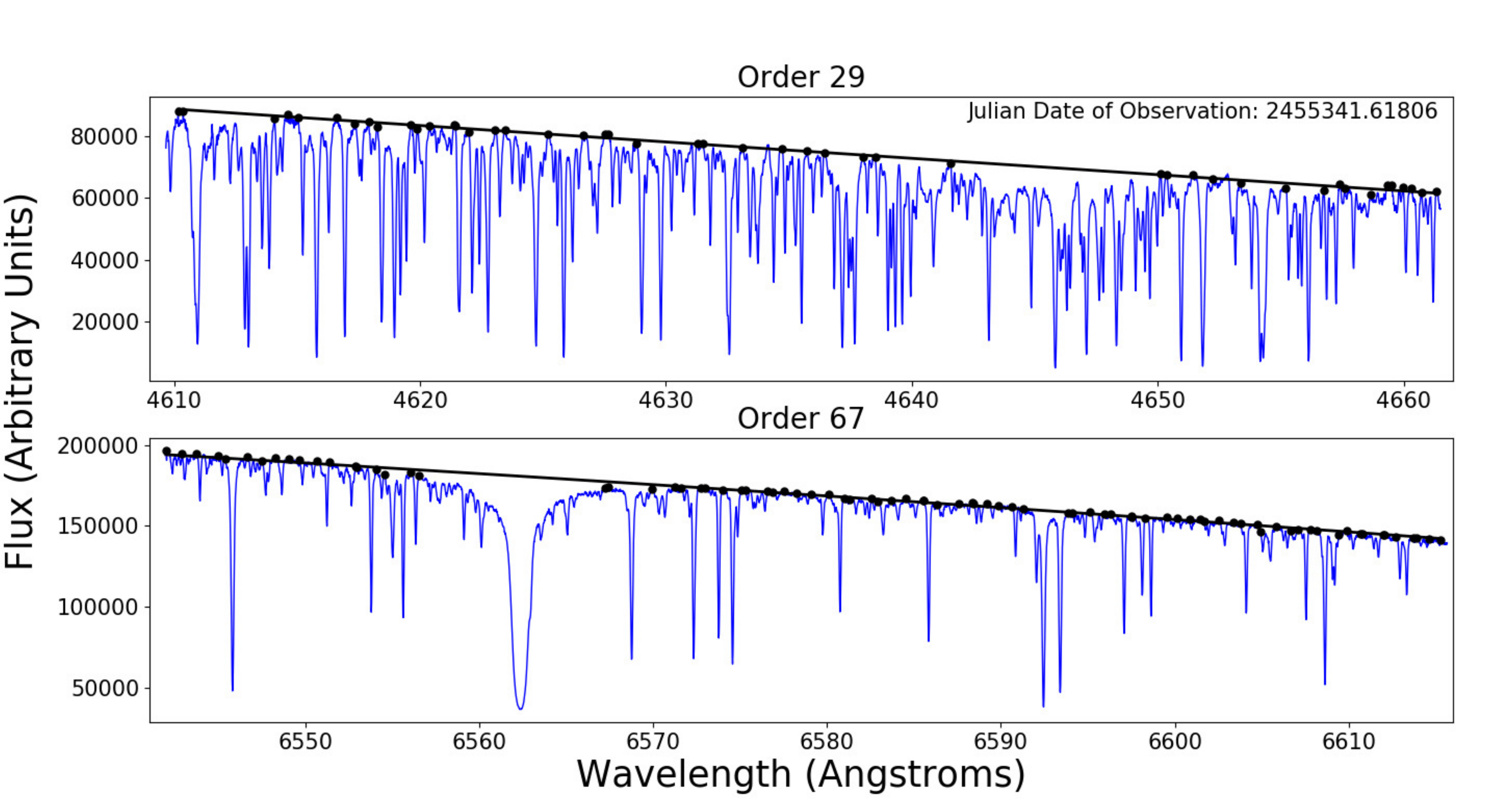}
\caption{Two example continuum fits, one for a blue order (number 29) and one for a red order (number 67) using a single $\alpha$~Cen~B spectrum. The original spectrum is shown in blue, the 2nd order polynomial fits to the continuum in black lines, and the unmasked LMPs used to generate the final fits in black circles. \label{fig10}}
\end{figure}

We have found our IME algorithm to generate continuum fits faster and more accurately than similar polynomial fitting algorithms that require manually selecting masked regions of the spectrum. Two example fits are shown in Figure~\ref{fig10}. Our algorithm contains many adjustable parameters, so we summarize our parameter choices for HARPS spectra in Table~\ref{params}. Similar applications of IME may be useful in any case of heteroscedastic noise when the regression function has significantly fewer degrees of freedom than the number of data points, and the heteroscedasticity cannot be easily modeled. When heteroscedastic noise is spatially sparse, such as in the redder orders of the spectrum where there are fewer spectral lines, IME performs particularly well.

\begin{table}
%\centering
\caption{Parameter Values\label{params}}
\begin{tabular}{ll}
\tableline\tableline
Parameter & Value\\
\tableline
Number of LMPs &100\\
Polynomial Fit Order &2\\
Initial Fit Tolerance &0.005\\
Minimum Fitted Points Per Side &4\\
Cosmic Ray Allowance &2\\
\tableline
& \\
\end{tabular}
%\tablenotetext{a}{See \cite{ss73}}
\end{table}

%\section{Combining normalized orders into a normalized 1-D spectrum}\label{orders}

\end{document}